\documentclass[prc,aps,twocolumn,showpacs]{revtex4}
\usepackage{graphicx}
\usepackage{amsbsy}
\begin{document}
\setlength{\baselineskip} {2.5ex}

\title{
 Measurement of $\pi^0 \Lambda$, $\bar{K}^0 n$,
 and $\pi^0 \Sigma^0$ production in $K^- p$ interactions
 for $p_{K^-}$ between 514 and 750~MeV/$c$
}

\author{
S.~Prakhov$^1$,
B.~M.~K.~Nefkens$^1$,
V.~Bekrenev$^2$, 
W.~J.~Briscoe$^3$,
N.~Knecht$^4$\footnote[1]{Present address: Physics Dept. of
   University of Toronto, 60 St.George St., Toronto, Ontario,
   Canada, M5S 1A7} 
A.~Koulbardis$^2$, N.~Kozlenko$^2$,
S.~Kruglov$^2$, 
G.~Lolos$^4$, I.~Lopatin$^2$,
A.~Maru\v{s}i\'{c}$^1$\footnote[2]{Present address: Collider-Accelerator 
Dept., Brookhaven National Laboratory, Upton, NY 11973, USA.},
S.~McDonald$^1$\footnote[3]{Present address: TRIUMF, 4004 Wesbrook Mall,
 Vancouver, B.C., Canada, V6T 2A3.},
D.~Peaslee$^5$\footnote[4]{Deceased},
 N.~Phaisangittisakul$^1$, 
J.~W.~Price$^1$,
A.~Shafi$^3$, 
A.~Starostin$^1$, H.~M.~Staudenmaier$^6$,
I.~I.~Strakovsky$^3$, and I.~Supek$^7$
}

\affiliation{
$^1$University of California Los Angeles, Los Angeles,
 California 90095-1547, USA}
\affiliation{
 $^2$Petersburg Nuclear Physics Institute, Gatchina 188350, Russia}
\affiliation{
 $^3$The George Washington University, Washington, D.C. 20052-0001, USA}
\affiliation{
 $^4$University of Regina, Saskatchewan, Canada, S4S OA2}
\affiliation{
 $^5$University of Maryland, College Park, Maryland 20742-4111, USA}
\affiliation{
 $^6$Universit\"at Karlsruhe, Karlsruhe 76128, Germany}
\affiliation{
 $^7$Rudjer Boskovic Institute, 10000 Zagreb, Croatia
}

\date{\today}
         
\begin{abstract}
 Differential cross sections and hyperon
 polarizations have been measured
 for $\bar{K}^0 n$, $\pi^0 \Lambda$,
 and $\pi^0 \Sigma^0$ production in $K^- p$ interactions
 at eight $K^-$ momenta between 514 and 750~MeV/$c$.
 The experiment detected the multiphoton final
 states with the Crystal Ball spectrometer using
 a $K^-$ beam from the Alternating Gradient Synchrotron
 of BNL.
 The results provide significantly greater precision than
 the existing data, allowing a detailed reexamination
 of the excited hyperon states in our energy
 range.
\end{abstract}


\maketitle

\section{Introduction}
 SU(3) Flavor Symmetry, FS, is a feature of QCD in the limit
 of vanishing quark masses. It implies many spectacular
 relations in baryon spectroscopy.
 For instance, it relates $d\sigma(K^- p \to \eta \Lambda)$
 to $d\sigma(\pi^- p \to \eta n)$ and 
 $d\sigma(K^- p \to \pi^0\pi^0\Lambda)$ to
 $d\sigma(\pi^- p \to \pi^0\pi^0 n)$.
 It also relates the width of flavor-symmetric states,
 such as $\Sigma(1385)\frac{3}{2}^+$ and
 $\Delta(1232)\frac{3}{2}^+$.
 FS implies that, for every three-quark $N^*$ resonance,
 there exists its flavor-symmetric $\Lambda^*$ state,
 which should have the same spin and parity as the $N^*$
 and have the flavor structure of SU(3) octet members.
 The mass of the $\Lambda^*$ resonance should be larger than that
 of the $N^*$ by the constituent $s-d$ quark-mass difference,
 which is about 140~MeV.
 Furthermore, there could exist also a $\Lambda^*$ that
 is a SU(3) flavor singlet. However, the $\Lambda^*$ singlet
 does not exist because it is not allowed as a consequence
 of color symmetry. FS also predicts that a triplet
 of $\Sigma^*$ states should exist for every $N^*$ octet and
 $\Delta^*$ decuplet.      
 
 The Review of Particle Properties~\cite{PDG} lists
 several candidates for $\Sigma^*$ states
 in the energy range between 1.5 and 1.7~GeV,
 the parameters of which are not well-established,
 and their status is still controversial.
 Three of these states, the one-star $\Sigma(1480)$ and 
 the two-star $\Sigma(1560)$ with unknown $J^P$ and 
 the one-star $\Sigma(1580)\frac{3}{2}^-$, cannot
 fulfill the requirements of FS and are therefore
 candidates for being either exotic states, such as
 five-quark states, or a hybrid state with an important
 gluon component in the wave function.  
 This situation is a consequence of the lack
 of reliable data on $K^- p$
 interactions~\cite{Arm70,Alston_tot,Alston,London75},
 especially for the reactions
 that have several neutral particles in the final state.
 
 The experimental study of the reactions $K^- p\to neutrals$ was
 performed with the Crystal Ball (CB) multiphoton spectrometer
 at the Alternating Gradient Synchrotron (AGS), aiming at
 a substantial improvement of the hyperon-spectroscopy field.  
 The major interest was in the production
 of the $\eta \Lambda$, $\bar{K}^0 n$, $\pi^0\Lambda$, $\pi^0\Sigma^0$,
 $\pi^0\pi^0\Lambda$, and $\pi^0\pi^0\Sigma^0$ final states.
 In these reactions, the experimental situation was especially
 poor for the $\pi^0\Sigma^0(\to \pi^0\gamma\Lambda)$,
 $\pi^0\pi^0\Lambda$, and $\pi^0\pi^0\Sigma^0$ production.
 These final states have more than one neutral particle
 and could not be measured in the old bubble-chamber experiments.
 Our results on the $\eta \Lambda$, $\pi^0\pi^0\Lambda$, and
 $\pi^0\pi^0\Sigma^0$ production in $K^- p$ interactions
 at eight incident $K^-$ momenta between 514 and 750~MeV/$c$
 have been reported in Refs.~\cite{etalam,l2pi0,s2pi0}.
 An analysis of our $K^- p \to \pi^0\Lambda$ data in the
 c.m. energy range 1565 to 1600~MeV to test the existence
 of the $\Sigma(1580)\frac{3}{2}^-$ state has been presented
 in Ref.~\cite{sig1580}.
 An independent analysis of the $K^- p \to \pi^0\Sigma^0$
 reaction, using the same data set,
 was recently presented in Ref.~\cite{SPi0}.
 Another independent analysis of the $K^- p \to \bar{K}^0 n$
 and $K^- p \to \pi^0\Lambda$ reactions was presented
 in Ref.~\cite{OlmPhD}.
 In the present work, we report the experimental
 results on $\bar{K}^0 n$, $\pi^0\Lambda$, and $\pi^0\Sigma^0$
 production in $K^- p$ interactions
 at eight incident $K^-$ momenta between 514 and 750~MeV/$c$.
 Compared to the analyses of Refs.~\cite{SPi0,OlmPhD},
 our analysis uses kinematic
 fitting as a part of the event reconstruction.
 The subtraction of background reactions
 is based on our own measurement of them.
\begin{figure*}
\includegraphics[width=10.cm,height=9.cm,bbllx=0.5cm,bblly=0.5cm,bburx=16.5cm,bbury=16.cm]{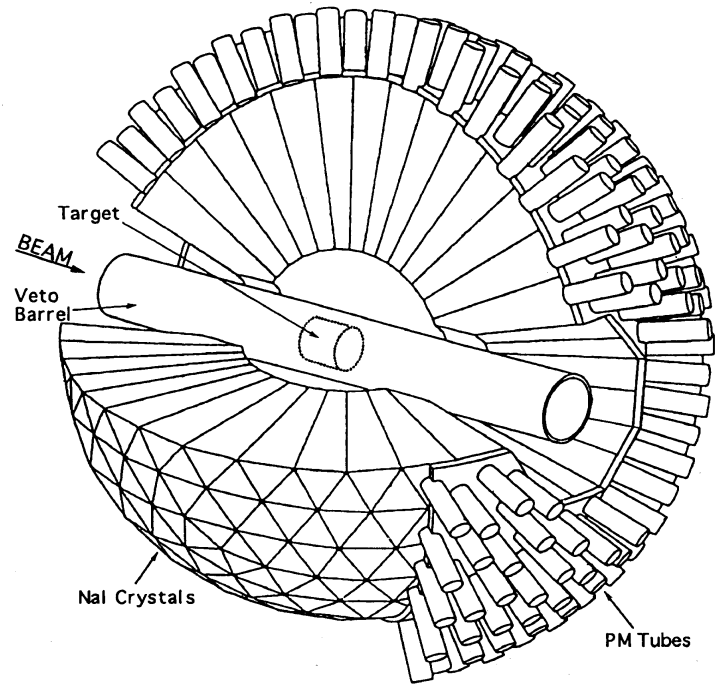}
\caption{
 Layout of the Crystal Ball spectrometer. 
}
 \label{fig:cball} 
\end{figure*}
 In contrast to Ref.~\cite{SPi0}, we use the full
 fiducial volume of the CB spectrometer, which provides us
 with much larger statistics and non-zero acceptance for
 the forward production angles.
 A new Partial Wave Analysis (PWA), involving our results on
 the differential cross sections
 for $\bar{K}^0 n$, $\pi^0 \Lambda$, 
 and $\pi^0 \Sigma^0$ production in $K^- p$ interactions
 and the polarizations of $\Lambda$ and $\Sigma^0$,
 is planned. This PWA is expected to improve the parameters
 of the low-mass $\Lambda^*$ and $\Sigma^*$ states.

\section{Experimental arrangements}

\begin{figure}
\includegraphics[width=5.cm,height=5.cm,bbllx=0.5cm,bblly=1.cm,bburx=15.cm,bbury=18.5cm]{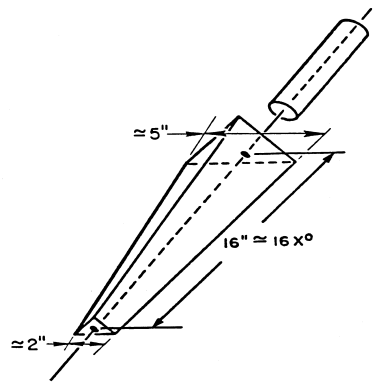}
\caption{
 Dimensions of the CB crystal. 
}
 \label{fig:cbcrystal} 
\end{figure}

 The Crystal Ball multiphoton spectrometer was installed
 in the C6 beam line of the AGS at Brookhaven National Laboratory.
 The CB spectrometer is a highly-segmented sphere
 of 672 NaI(Tl) crystals (see Fig.~\ref{fig:cball}).
 The sphere has an entrance and exit tunnel for the beam
 and a 50-cm-diameter spherical cavity in the center.
 The solid angle covered by the CB is 93\% of $4\pi$ steradian.
 The CB crystals are packed in two
 hermetically sealed and evacuated hemispheres.
 The crystals have the shape of truncated
 triangular pyramid (see Fig.~\ref{fig:cbcrystal}),
 all pointed towards the center of the CB.
 The crystal length is 40.6~cm,
 which corresponds to 15.7 radiation lengths.

 The pulse height in every crystal was measured with individual ADCs
 (analog-to-digital converter).
 For registering the timing information, one TDC
 (time-to-digital converter) was used for
 every minor triangle, which is a group of nine neighboring crystals. 
 The typical energy resolution for electromagnetic showers in the CB
 was $\Delta E/E = 0.020/(E[\mathrm{GeV}])^{0.36}$.
 The directions of the photon showers were measured with a resolution
 in $\theta$ (the polar angle with respect to the beam axis)
 of $\sigma_\theta = 2^\circ \textrm{---} 3^\circ$,
 under the assumption that the photons are
 produced in the center of the CB.
 The resolution in
 azimuthal angle $\phi$ is $\sigma_\theta/\sin\theta$. 
 The angular resolutions are mainly defined by the granularity
 of the CB and do not depend on the
 energy resolution of the CB crystals.
 The situation is slightly different when a photon
 is produced not in the center of the CB. In this case, the real
 angles of the photon must be calculated from an interaction point
 (which can be either the primary vertex of the event in the target
  or the secondary vertex of the outgoing particles decaying in flight)
  to the point inside the CB that is
 determined by the depth of the electromagnetic shower
 in the NaI(Tl) material.
 Also, the energy and angular resolution of the CB becomes somewhat
 worse when a photon hits the crystals close to
 the CB beam tunnels. In this case, part of the electromagnetic
 shower can leak through the side surface of the so-called
 ``guard'' crystals (i.e., the crystals that form the inner surface
 of the tunnels).

 The C6 line provided a beam of negative kaons and pions with
 the $K^-/\pi^-$ ratio enhanced to about 0.1
 by two electrostatic separators.
 Beam particles were incident on a 10-cm-long
 liquid-hydrogen (LH$_2$) target located in the center
 of the Crystal Ball. The beam divergences on the target
 were $\sigma_X \approx 2.5$~cm and $\sigma_Y \approx 2$~cm.
 The diameter of the target was 10~cm.
 The determination of the beam momentum was
 done with a dipole magnet and a set of four multiwire
 drift chambers. One of the chambers was located upstream
 and the other three downstream of the magnet. The coordinate
 resolution of the chambers was
 $\sigma_X \approx \sigma_Y \approx 0.2$~mm.
 The momentum resolution $\sigma_p/p$ for an individual
 incident kaon varied from 0.6\% to 1.\%, depending on the momentum value.
 The largest contribution to the uncertainty
 in the incident-particle momentum comes from the kaon
 multiple scattering and energy losses in the beam counters
 and the LH$_2$ target. The mean value $p_{K^-}$ for
 the incident-momentum spectra 
 and the momentum spread $\sigma_p$, which were determined
 at the target center, are listed in Table~\ref{tab:events}.
 The uncertainty in determining the mean beam momentum is 2---3~MeV/$c$.
 This uncertainty is mainly caused by the precision of the calculation
 of the magnetic field in the dipole magnet and the energy loss
 of the incident kaons before they interact in the target. 

 To monitor the beam particles, a system of plastic scintillation counters
 was used. This system included the $S_1$ hodoscope upstream
 and the $S_2$ counter downstream of the dipole magnet,
 and the $S_T$ counter
 located 162~cm upstream of the LH$_2$ target. The coincidence
 of signals from $S_1$, $S_2$ and $S_T$ served as the beam trigger.
 The time-of-flight of beam particles between $S_1$ and $S_T$
 was used in the trigger and also in the off-line analysis
 to suppress the background from pion interactions in the target.

 The LH$_2$ target was surrounded by a 16-cm-diameter ``barrel'' made
 of four  plastic scintillation counters that functioned as a veto
 for the beam interactions with charged particles in the final state.
 The 120-cm length of the veto-barrel counters ensured almost
 100\% rejection of those events. The 5-mm thickness
 of the counters implied both a good efficiency of the veto barrel
 for the charged-particle detection and a low probability
 for photon conversion. 

 The CB event trigger was
 the beam trigger in coincidence with a CB energy
 trigger, which required the total energy deposited
 in the crystals to exceed a certain threshold.
 The first two layers of the crystals in the CB tunnels
 (i.e., closest to the beam line),
 which had a high occupancy from beam-halo interactions,
 were excluded from the CB energy trigger.
 The CB neutral-event trigger 
 required the anti-coincidence of the CB event trigger
 with signals from the veto-barrel counters.

 More details about the CB setup at the AGS and its $K^- p$-data analyses
 can be found in Refs.~\cite{etalam,l2pi0,s2pi0,sig1580,SPi0,OlmPhD,3pi0lam,lam1670}.

\section{Data analysis}

\begin{figure}
\includegraphics[width=5.cm,height=5.cm,bbllx=2.5cm,bblly=2.cm,bburx=16.5cm,bbury=17.cm]{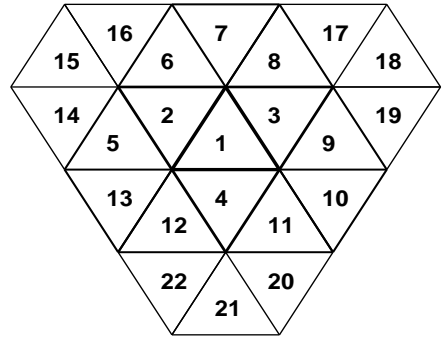}
\caption{
 Configuration of a cluster in the CB. 
}
 \label{fig:cbclsdraw} 
\end{figure}

 Since the Crystal Ball detector is designed as a
 multiphoton spectrometer, $K^- p$ interactions
 were studied by measuring the photons and the neutron
 in the final state.
 The $\bar{K}^0$ meson and the $\Lambda$ and
 $\Sigma^0$ hyperons were measured in the CB by
 the decay chains $K^0_S \to \pi^0\pi^0 \to 4\gamma$,
 $\Lambda \to \pi^0 n \to 2\gamma n$, and
 $\Sigma^0 \to \gamma\Lambda \to \gamma\pi^0 n \to 3\gamma n$.
 As the final-state photons produce electromagnetic showers in the NaI(Tl)
 crystals, they can be recognized as so-called ``clusters'' in the density
 of the energy deposited in the CB.
 The outgoing neutrons can also be detected if the products
 of their interactions in the NaI(Tl) material produce
 the ionization that is enough to form a cluster. 
 In general, a cluster in the CB is defined as
 a group of neighboring crystals in which energy is 
 deposited from the interaction of a photon,
 a charged particle, or a neutron produced in the
 final state. Clusters produced by different particle types 
 have different features.
 Since our analysis involves mainly
 multi-photon final states, the cluster algorithm was
 optimized to reconstruct parameters of a photon from
 its electromagnetic shower with good energy and angular resolution
 and with a reasonable separation between shower split-offs
 and overlapping showers.
 The software threshold for the cluster energy was chosen to be 20~MeV;
 this value optimizes the yield of the
 reconstructed events for the $K^- p\to neutrals$ processes under
 study.
 The size of one cluster was limited to a configuration
 of 22 crystals (see Fig.~\ref{fig:cbclsdraw}), which is
 large enough to contain the full spread of
 a photon shower at our energies. 
 Cluster directions were measured as the weighted average
 of the directions of all crystals forming the cluster.
 The weight factor of each crystal in this average
 was taken as the square root of the crystal energy, $\sqrt{E}$. 
\begin{figure*}
\includegraphics[width=15.5cm,height=5.5cm,bbllx=0.5cm,bblly=0.5cm,bburx=19.5cm,bbury=7.cm]{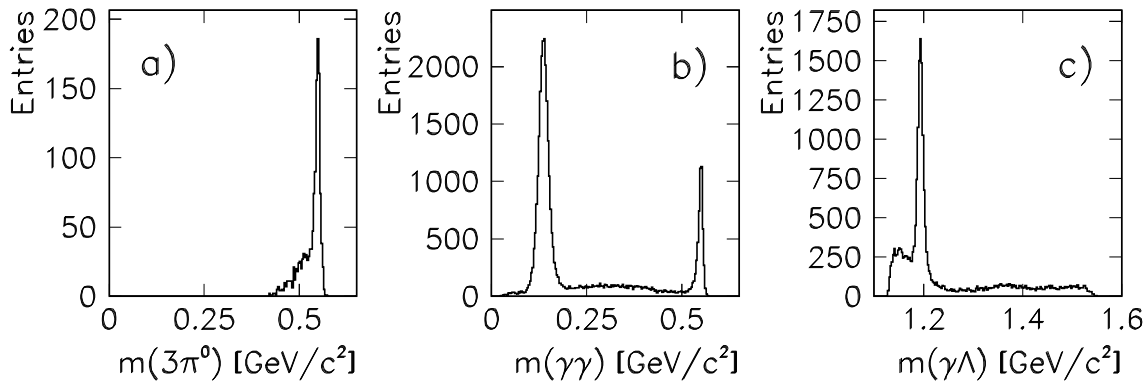}
\caption{
 The invariant-mass spectra for the experimental events
 at $p_{K^-} = 750$~MeV/$c$
 selected by testing the following hypotheses with the kinematic fit:
 (a) $K^-p \to \pi^0\pi^0\pi^0\Lambda \to 4\pi^0 n \to 8\gamma n$,
  where $\sigma_m(\eta\to 3\pi^0) \approx 5$~MeV/$c^2$;
 (b) $K^-p \to \gamma\gamma\Lambda \to \gamma\gamma\pi^0 n \to 4\gamma n$,
  where $\sigma_m(\eta\to 2\gamma) \approx 6$~MeV/$c^2$ and
  $\sigma_m(\pi^0\to 2\gamma) \approx 13$~MeV/$c^2$;
 and (c) $K^-p\to \gamma\pi^0\Lambda\to \gamma\pi^0\pi^0 n\to 5\gamma n$,
 where $\sigma_m(\Sigma^0\to \gamma\Lambda) \approx 6$~MeV/$c^2$.  
}
 \label{fig:mass_resl} 
\end{figure*}

 As the decay time of NaI(Tl) is about 250~ns, the high intensity
 of the incident beam causes
 good events to be contaminated by so-called pile-up clusters
 (i.e., the clusters remaining from other events).
 Since the pile-up clusters change the proper number
 of the clusters expected in the CB from good events, such
 events can be lost from consideration
 unless the pile-up clusters are eliminated.
 Elimination of the pile-up clusters was based on the
 TDC information of the crystals
 forming the clusters.
 Since our good events cause the CB trigger, their clusters
 have times that are peaked in a restricted timing window. 
 All clusters that occur outside this
 window were eliminated from further consideration.
 The loss of good events due to the pile-up clusters
 inside the window
 were estimated by varying the window width.
 
 The kinematic-fitting technique was used to select
 candidates for the reactions that were studied.
 In the present analysis, every event with
 the proper cluster multiplicity was fitted to the following
 three hypotheses:
\begin{equation}
   K^- p \to K^0_S n \to (\pi^0 \pi^0) n \to 4\gamma n~,
\label{eqn:k0sn}
\end {equation} 
\begin{equation}
   K^- p \to \pi^0 \Lambda \to \pi^0 (\pi^0 n) \to 4\gamma n~,
\label{eqn:lpi0}
\end{equation}
\begin{equation}
 K^-p\to \pi^0\Sigma^0\to \pi^0\gamma\Lambda\to \pi^0\pi^0\gamma n \to 5\gamma n~.
 \label{eqn:spi0}
\end{equation}
 The incident kaon was parameterized in the kinematic fit
 by five measured variables: momentum,
 angles $\theta_x$ and $\theta_y$, and position coordinates
 $x$ and $y$ at the target.
 A photon cluster was parameterized by three measured variables:
 energy and angles $\theta$ and $\phi$. 
 As the data were taken with a 10-cm-long LH$_2$
 target, the $z$ coordinate of the vertex was a free variable
 of the kinematic fit. Including $z$ into the fit
 improves the angular resolution of
 the photons. The cluster angles
 $\theta$ and $\phi$, which were used in the minimization
 procedure, were calculated with respect to the CB center.
 The photon angles, used in the calculation of
 the kinematic constraints, were defined by
 the directions from the vertex coordinates
 to the point determined by the cluster angles and
 the effective depth of the photon shower in the NaI material. 
 In our analysis, the effective depth of the electromagnetic
 shower was defined as the depth where a photon has likely deposited
 half its initial energy. For the neutron cluster,
 the effective depth was taken to be half the length of the crystals.

 When the final-state neutron was not detected, its
 energy and two angles were free variables of
 the kinematic fit. For the neutron detected in the CB, the angles of its cluster
 were used as the measured variables and the neutron energy as
 a free variable.
 In our analysis of reactions~(\ref{eqn:k0sn}), (\ref{eqn:lpi0}),
 and (\ref{eqn:spi0}), the detection efficiency for the final-state
 neutron increases from 21\% for our lowest beam momenta
 to 28\% for the highest.
 Since the major ionization from the neutron passage in the crystals
 occurs due to recoil protons, the energy of the neutron
 cluster is very uncertain and cannot be used to reconstruct the
 kinetic energy of the neutron. Many neutrons, especially low-energy
 ones, produce a signal in the CB that is below the 20-MeV-energy
 threshold used in the cluster reconstruction.
 The angular resolution of the CB for neutron clusters
 is slightly worse than for photon clusters.

 Since all reactions in our analysis have a particle decaying
 in flight, the decay length of this particle was also
 a free variable of the kinematic fit.
 The corresponding secondary vertex was then
 determined from the primary-vertex coordinates, the direction
 of the decaying particle, and the decay length.
 The calculation of the angles of the final-state photons and neutron
 from the cluster angles, if they were
 produced from the secondary vertex, is the same as for
 the primary vertex.
\begin{figure*}
\includegraphics[width=17.cm,height=9.cm,bbllx=1.cm,bblly=1.cm,bburx=19.5cm,bbury=11.cm]{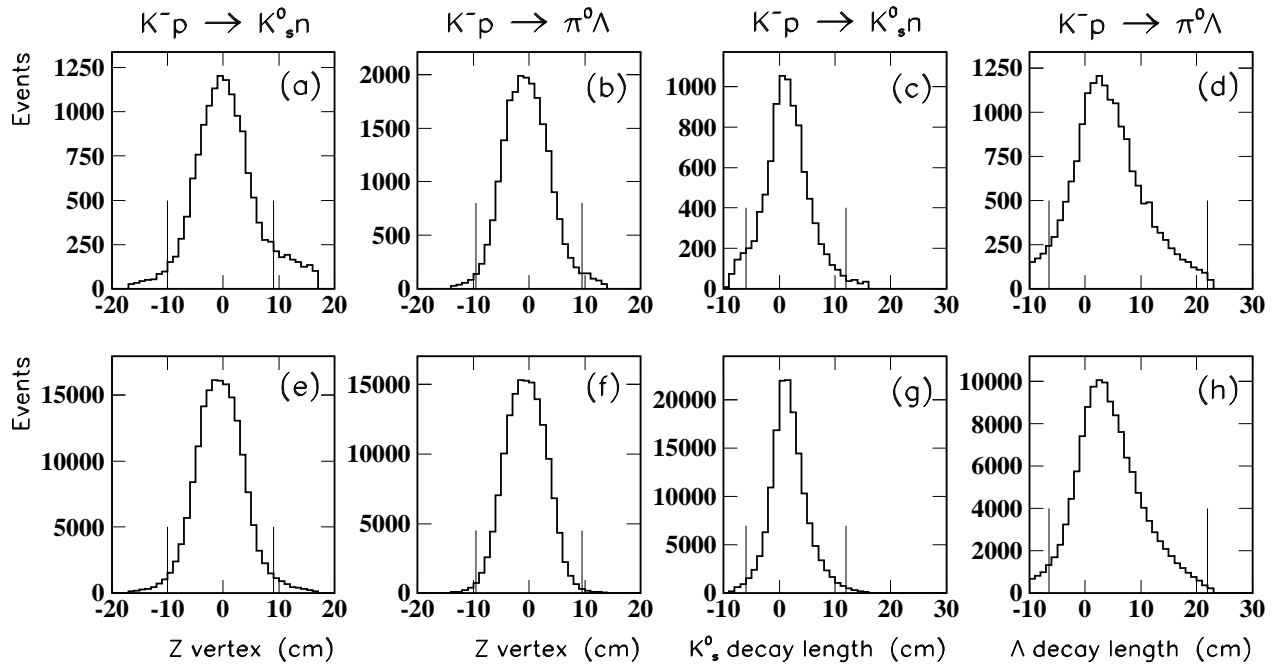}
\caption{
 Experimental (a-d) and MC (e-h) distributions for the $z$ coordinate
 of the primary vertex
 in reaction (a,e) $K^-p \to K^0_S n$ and (b,f) $K^-p \to \pi^0 \Lambda$,
 and for the decay length of 
 (c,g) $K^0_S$ in reaction $K^-p \to K^0_S n$ and
 (d,h) $\Lambda$ in reaction $K^-p \to \pi^0 \Lambda$.
 The events between the vertical lines shown in the figures are accepted
 for further analysis.
}
 \label{fig:zvrt_dlen_kn_lp} 
\end{figure*}

 The invariant-mass resolution of the CB after applying
 the kinematic-fitting technique is illustrated
 in Fig.~\ref{fig:mass_resl}.
 Three invariant-mass spectra are shown for our data
 at the beam momentum of 750~MeV/$c$ after testing events for
 the following hypotheses:
 $K^-p \to \pi^0\pi^0\pi^0\Lambda$, $K^-p \to \gamma\gamma\Lambda$,
 and $K^-p \to \pi^0\gamma\Lambda$,  
 where the $\Lambda$ hyperon was identified by its decay into $\pi^0 n$.
 The invariant-mass resolution for the $m(3\pi^0)$ peak
 from $\eta$ decays has $\sigma_m(3\pi^0) \approx 5$~MeV/$c^2$,
 which is comparable to $\sigma_m(\gamma\gamma) \approx 6$~MeV/$c^2$
 from $\eta\to\gamma\gamma$ decays.
 The $m(\gamma\gamma)$ resolution for $\pi^0\to\gamma\gamma$ decays
 is about 13~MeV/$c^2$.
 We cannot illustrate the $m(\pi^0 n)$ resolution for
 $\Lambda \to \pi^0 n$ decays
 as a constraint on the $\Lambda$-hyperon mass is used in the 
 kinematic fit for the secondary-vertex determination.
 Instead, we show the $m(\gamma\Lambda)$ peak from $\Sigma^0$ decays,
 for which $\sigma_m(\gamma\Lambda) \approx 6$~MeV/$c^2$. 

 The candidates for reactions~(\ref{eqn:k0sn}) and (\ref{eqn:lpi0})
 were searched for in four-cluster and five-cluster events
 since these processes have the same final state,
 including four photons and the neutron. 
 The four-cluster events were tested for
 the case when only the four photons were detected in the CB.
 The five-cluster events were tested for the case when all five
 final-state particles were detected. Similarly, the candidates for
 reaction~(\ref{eqn:spi0}), which have five photons and the neutron
 in the final state, were searched for in five-cluster and six-cluster
 events. The test of each hypothesis involves all possible permutations
 of assigning the detected clusters to the particles in the reaction chain.   
 The events for which at least one permutation satisfied
 the tested hypothesis at the 2\% confidence level, C.L.,
 (i.e., with a probability larger than 2\%)
 were accepted as the reaction candidates.
 The permutation with the largest C.L. was used
 to reconstruct the kinematics of the reaction.
\begin{figure*}
\includegraphics[width=15.0cm,height=10.cm,bbllx=0.5cm,bblly=1.0cm,bburx=19.5cm,bbury=14.cm]{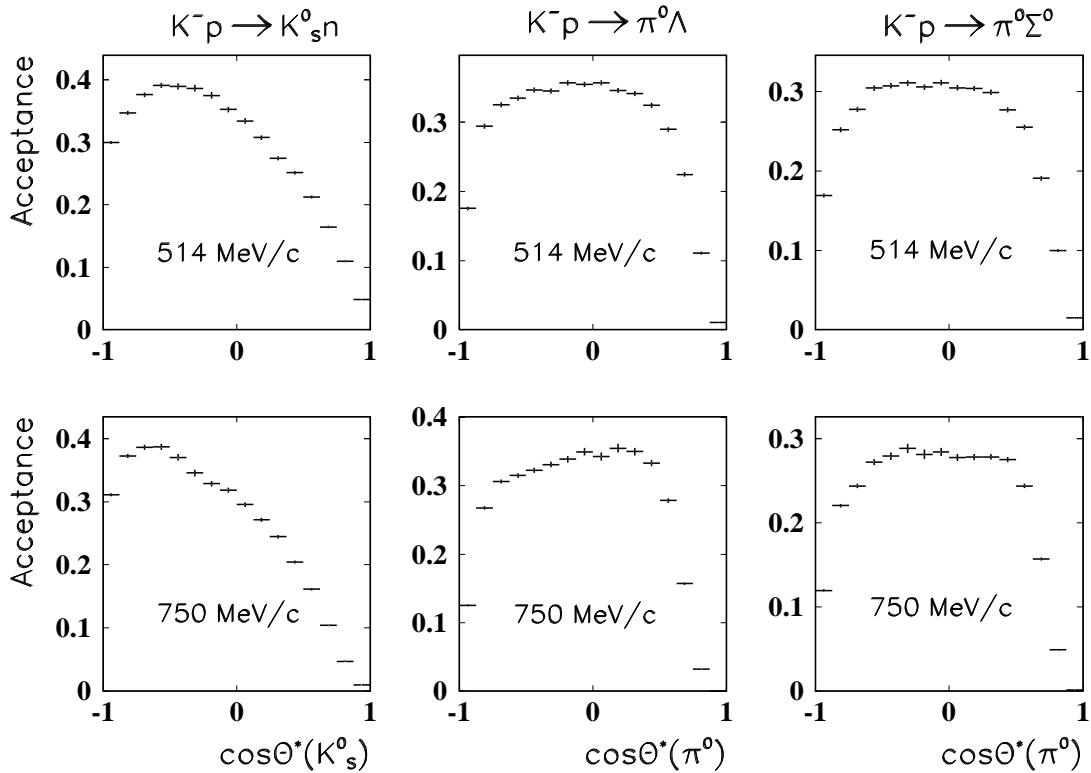}
\caption{
 The acceptance for the production angle $\theta^*$ of the outgoing meson
 in the c.m. system; it is shown for reactions $K^-p \to K^0_S n$,
 $K^-p \to \pi^0 \Lambda$, and $K^-p \to \pi^0 \Sigma^0$ at beam
 momenta of 514 and 750~MeV/$c$.  
}
 \label{fig:k0sn_lpi0_spi0_dxs_acc} 
\end{figure*}

 The candidates selected for our three reactions are contaminated
 with backgrounds that must be subtracted in the analysis.
 The first source of background is from
 misidentification of events from different $K^-p$ reactions.
 The second source is from processes that are not kaon interactions
 in the LH$_2$ target. The major fraction of these interactions are
 $K^-$ decays in the beam. This background was investigated
 using data samples when the target was empty.
 The absolute contribution of this background was
 determined from the ratio of totals of the beam kaons incident on
 the full and empty target. The fraction of the so-called ``empty-target''
 background, which left in the reaction candidates after all selection cuts,
 varies slightly depending on the beam momentum, the reaction itself,
 and the cuts applied. For our typical selection cuts, which are discussed
 later on in the text, the remaining empty-target background
 comprises 6\% to 9\% for the $K^0_S n$ and $\pi^0\Lambda$ candidates
 and 4\% to 6\% for the $\pi^0\Sigma^0$ candidates.
 This background was subtracted from our experimental production-angle
 distributions.
 
 A Monte Carlo (MC) simulation of reactions~(\ref{eqn:k0sn}), (\ref{eqn:lpi0}),
 and (\ref{eqn:spi0}) was used to determine the experimental acceptance
 and to estimate the fraction of the background coming from
 misidentification of events from different reactions.
 First, all reactions were simulated with an isotropic
 production-angle distribution.
 To reproduce the beam structure correctly, the information from
 the experimental beam-trigger events was used as the input for simulating
 the incident kaons. The MC events for each beam momentum were then 
 propagated through a full {\sc GEANT} (version 3.21)
 simulation of the CB detector, folded with the CB resolutions and
 trigger conditions, and analyzed the same way as the experimental data.
 The small difference between the experimental data
 and the MC simulation for the neutron response
 in the CB was not important, 
 as we summed the events with and without the neutron detected.

 The analysis of the MC simulation showed that the largest misidentification
 of events from different reactions is between processes~(\ref{eqn:k0sn})
 and (\ref{eqn:lpi0}), which have the same final state of four photons.
 However, for the events that satisfied both the hypotheses,
 the C.L. for the ``true'' one was typically larger than for the ``false''.
 Therefore, to decrease the background of these reactions
 from each other to the level of 4\% or less,  
 we accepted only the hypothesis with the largest C.L., also requiring it
 to be at least twice as large as the C.L. for the second hypothesis.
 For further suppression of other backgrounds, tightening the cut on the C.L.
 of the reaction hypothesis itself can be applied.
 If the events of each the reaction are selected at the 5\% C.L.,
 the background from process~(\ref{eqn:spi0}) was found to be less
 than 1\% for $K^0_S n$ events and less than 2\% for $\pi^0\Lambda$
 events. The $\pi^0\Sigma^0$ events were contaminated
 by a background from process~(\ref{eqn:lpi0}) to the level less than 2\%, 
 and from processes~(\ref{eqn:k0sn}) and
 $K^- p \to \pi^0 \pi^0 \Lambda \to 3\pi^0 n$ to the level less than 1\%.

 The low levels of the background contamination
 were achieved in part by applying a cut on the $z$ coordinate of the primary
 vertex of the event and on the decay length of 
 the $K^0_S$ meson and the $\Lambda$ hyperon.
 The $z$-coordinate and decay-length distributions together with
 the cuts are illustrated for the experimental data
 and the MC simulation in
 Fig.~\ref{fig:zvrt_dlen_kn_lp}.
 The small difference between the experimental data
 and MC simulation is due to the remaining background
 that is not subtracted from the experimental distributions.
 The deviation of the $z$ distributions from the target shape is determined
 by the resolution of the kinematic fit in $z$, which is about 2~cm. 
 Similarly, the actual decay-length distributions are smeared with the resolution
 of 3~cm.

 The acceptance for reactions $K^-p \to K^0_S n$,
 $K^-p \to \pi^0 \Lambda$, and $K^-p \to \pi^0 \Sigma^0$ was
 determined as a function of $\cos(\theta^*)$,
 where $\theta^*$ is the production angle of
 the final-state meson (i.e., the angle between the direction
 of the outgoing meson and the incident $K^-$ meson) 
 in the center-of-mass (c.m.) system.
 The acceptance for each of the studied reactions is
 shown in Fig.~\ref{fig:k0sn_lpi0_spi0_dxs_acc} for two beam momenta:
 514 and 750~MeV/$c$.
 These acceptances include the effects of all
 our standard cuts used for the event selection. 
 A poorer $\cos(\theta^*)$ acceptance of the forward $\theta^*$ angles for a higher
 beam momentum is mostly due to the larger threshold on the CB total energy
 in the event trigger. This threshold was 0.9~GeV for $p_{K^-} = 514$~MeV/$c$
 and reached 1.5~GeV for $p_{K^-} = 750$~MeV/$c$.
 The experimental resolution in $\theta^*$ for $K^-p \to K^0_S n$ varies between
 $6.5^\circ$ at $p_{K^-} = 514$~MeV/$c$ and $4.5^\circ$ at $p_{K^-} = 750$~MeV/$c$.
 The $\theta^*$ resolution for $K^-p \to \pi^0 \Lambda$ and $K^-p \to \pi^0 \Sigma^0$
 varies between $3.5^\circ$ at $p_{K^-} = 514$~MeV/$c$ and
 $3.0^\circ$ at $p_{K^-} = 750$~MeV/$c$.   

 The experimental $\cos(\theta^*)$ distributions, which were obtained after
 the ``empty-target'' subtraction and the acceptance correction, were used
 to simulate reactions $K^-p \to K^0_S n$,
 $K^-p \to \pi^0 \Lambda$, and $K^-p \to \pi^0 \Sigma^0$ with the realistic
 production-angle distributions. Then this MC simulation was used for
 the subtraction of the background remaining from the reaction
 misidentification.
\begin{table*}
\caption
[tab:events]{
 Number of experimental events obtained for reactions
 $K^- p \to K^0_S n$, $K^- p \to \pi^0 \Lambda$, and 
 $K^- p \to \pi^0 \Sigma^0$ at the eight beam momenta.
 } \label{tab:events}
\begin{ruledtabular}
\begin{tabular}{|l|c|c|c|c|c|c|c|c|} 
\hline
 $p_{K^-}\pm \sigma_p$ [MeV/$c$] &
 514$\pm$10 & 560$\pm$11 & 581$\pm$12 & 629$\pm$11 &
 659$\pm$12 & 687$\pm$11 & 714$\pm$11 & 750$\pm$13 \\
\hline
 $N_{\mathrm{Exp}}(K^- p \to K^0_S n)$ & 2,514 & 5,007 & 6,968 & 7,711 
                          & 7,903 & 8,245 & 8.157 & 10,134 \\
 $N_{\mathrm{Exp}}(K^- p \to \pi^0 \Lambda)$ & 3,402 & 6,689 & 10,118 & 11,735 
                           & 13,238 & 15,157 & 16,510 & 25,948 \\
 $N_{\mathrm{Exp}}(K^- p \to \pi^0 \Sigma^0)$ & 2,702 & 5,041 & 7,269 & 7,484 
                           & 8,092 & 8,964 & 9,684 & 14,010 \\
\hline
\end{tabular}
\end{ruledtabular}
\end{table*}
\begin{figure*}
\includegraphics[width=16.cm,height=15.cm,bbllx=0.5cm,bblly=1.cm,bburx=19.5cm,bbury=18.5cm]{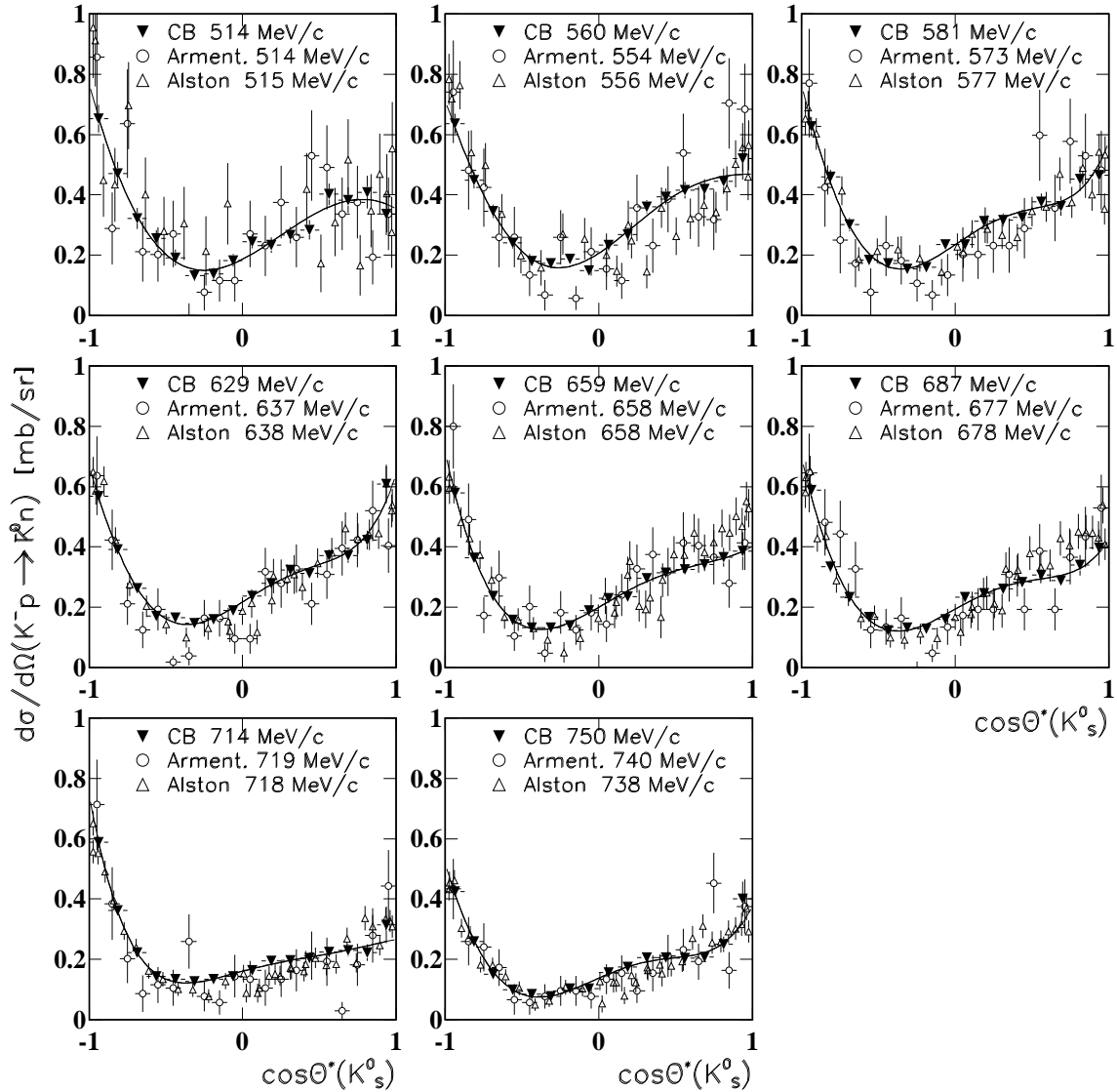}
\caption{
 Our differential cross sections for $K^-p \to \bar{K}^0 n$ compared
 to the results of Armenteros~\protect\cite{Arm70} and
 Alston-Garnjost~\protect\cite{Alston} for similar beam momenta. 
 The curves are the Legendre polynomial fits of our data. 
}
 \label{fig:k0sn_dxs_v13_arm_als} 
\end{figure*}
 The subtraction of the $\pi^0 \pi^0 \Lambda$ background from the
 $\pi^0 \Sigma^0$ spectra was based on the results of our analysis
 for $K^-p \to \pi^0 \pi^0 \Lambda$ published in Ref.~\cite{l2pi0}.
 
\section{Experimental results}

 The numbers of experimental events remaining
 after the subtraction of all backgrounds are listed for our three
 reactions and the eight beam momenta in Table~\ref{tab:events}. 
\begin{table*}
\caption
[tab:k0sn1]{
 Differential cross sections for the $K^- p \to {\bar K}^0 n$ reaction
 for the four lowest beam momenta.
 } \label{tab:k0sn1}
\begin{ruledtabular}
\begin{tabular}{|c|c|c|c|c|} 
\hline
 $p_{K^-}\pm \delta_p$ [MeV/$c$] & $514\pm 10$ & $560\pm 11$ &
                                     $581\pm 12$ & $629\pm 11$ \\
\hline
  $\cos\theta^{\ast}({\bar K}^0)$ & $d\sigma/d\Omega$ [mb/sr] &
  $d\sigma/d\Omega$ [mb/sr]&  $d\sigma/d\Omega$ [mb/sr] &  $d\sigma/d\Omega$ [mb/sr]\\
\hline
 -0.94  & $0.653\pm 0.046$ & $0.636\pm 0.033$ & $0.626\pm 0.032$ & $0.568\pm0.022$ \\
 -0.81  & $0.471\pm 0.039$ & $0.449\pm 0.028$ & $0.459\pm 0.025$ & $0.391\pm0.017$ \\
 -0.69  & $0.322\pm 0.035$ & $0.346\pm 0.022$ & $0.301\pm 0.022$ & $0.263\pm0.014$ \\
 -0.56  & $0.255\pm 0.028$ & $0.241\pm 0.018$ & $0.183\pm 0.018$ & $0.171\pm0.011$ \\
 -0.44  & $0.192\pm 0.027$ & $0.181\pm 0.016$ & $0.173\pm 0.013$ & $0.165\pm0.010$ \\
 -0.31  & $0.133\pm 0.019$ & $0.172\pm 0.015$ & $0.155\pm 0.013$ & $0.147\pm0.009$ \\
 -0.19  & $0.138\pm 0.020$ & $0.188\pm 0.015$ & $0.159\pm 0.017$ & $0.160\pm0.010$ \\
 -0.06  & $0.182\pm 0.021$ & $0.149\pm 0.015$ & $0.235\pm 0.015$ & $0.190\pm0.011$ \\
 0.06  & $0.245\pm 0.023$ & $0.234\pm 0.018$ & $0.236\pm 0.018$ & $0.237\pm0.013$ \\
 0.19  & $0.233\pm 0.026$ & $0.269\pm 0.019$ & $0.313\pm 0.019$ & $0.279\pm0.014$ \\
 0.31  & $0.267\pm 0.027$ & $0.360\pm 0.023$ & $0.317\pm 0.022$ & $0.323\pm0.016$ \\
 0.44  & $0.284\pm 0.033$ & $0.394\pm 0.027$ & $0.325\pm 0.026$ & $0.313\pm0.017$ \\
 0.56  & $0.402\pm 0.037$ & $0.415\pm 0.028$ & $0.377\pm 0.027$ & $0.371\pm0.020$ \\
 0.69  & $0.382\pm 0.042$ & $0.418\pm 0.034$ & $0.363\pm 0.032$ & $0.372\pm0.025$ \\
 0.81  & $0.408\pm 0.054$ & $0.444\pm 0.043$ & $0.453\pm 0.041$ & $0.424\pm0.032$ \\
 0.94  & $0.337\pm 0.069$ & $0.521\pm 0.071$ & $0.466\pm 0.063$ & $0.609\pm0.064$ \\
\hline
\end{tabular}
\end{ruledtabular}
\end{table*}
\begin{table*}
\caption
[tab:k0sn2]{
 Differential cross sections for the $K^- p \to {\bar K}^0 n$ reaction
 for the four highest beam momenta.
 } \label{tab:k0sn2}
\begin{ruledtabular}
\begin{tabular}{|c|c|c|c|c|} 
\hline
 $p_{K^-}\pm \delta_p$ [MeV/$c$] & $659\pm 12$ & $687\pm 11$ &
                                     $714\pm 11$ & $750\pm 13$ \\
\hline
  $\cos\theta^{\ast}({\bar K}^0)$ & $d\sigma/d\Omega$ [mb/sr] &
  $d\sigma/d\Omega$ [mb/sr]&  $d\sigma/d\Omega$ [mb/sr] &  $d\sigma/d\Omega$ [mb/sr]\\
\hline
 -0.94  & $0.578\pm 0.020$ & $0.588\pm 0.022$ & $0.589\pm 0.019$ & $0.425\pm0.014$ \\
 -0.81  & $0.363\pm 0.017$ & $0.335\pm 0.019$ & $0.362\pm 0.014$ & $0.260\pm0.010$ \\
 -0.69  & $0.237\pm 0.013$ & $0.233\pm 0.015$ & $0.222\pm 0.011$ & $0.150\pm0.008$ \\
 -0.56  & $0.159\pm 0.010$ & $0.167\pm 0.010$ & $0.144\pm 0.008$ & $0.100\pm0.006$ \\
 -0.44  & $0.132\pm 0.010$ & $0.123\pm 0.011$ & $0.134\pm 0.008$ & $0.085\pm0.006$ \\
 -0.31  & $0.132\pm 0.010$ & $0.132\pm 0.009$ & $0.127\pm 0.009$ & $0.077\pm0.006$ \\
 -0.19  & $0.140\pm 0.010$ & $0.129\pm 0.011$ & $0.135\pm 0.008$ & $0.104\pm0.007$ \\
 -0.06  & $0.190\pm 0.011$ & $0.159\pm 0.011$ & $0.144\pm 0.009$ & $0.103\pm0.007$ \\
 0.06  & $0.229\pm 0.013$ & $0.233\pm 0.012$ & $0.165\pm 0.010$ & $0.156\pm0.008$ \\
 0.19  & $0.236\pm 0.014$ & $0.246\pm 0.012$ & $0.195\pm 0.011$ & $0.175\pm0.008$ \\
 0.31  & $0.296\pm 0.015$ & $0.261\pm 0.015$ & $0.196\pm 0.012$ & $0.207\pm0.010$ \\
 0.44  & $0.315\pm 0.017$ & $0.283\pm 0.016$ & $0.204\pm 0.013$ & $0.206\pm0.011$ \\
 0.56  & $0.326\pm 0.019$ & $0.307\pm 0.019$ & $0.225\pm 0.015$ & $0.204\pm0.012$ \\
 0.69  & $0.343\pm 0.023$ & $0.289\pm 0.019$ & $0.230\pm 0.020$ & $0.207\pm0.015$ \\
 0.81  & $0.366\pm 0.032$ & $0.339\pm 0.034$ & $0.222\pm 0.028$ & $0.251\pm0.025$ \\
 0.94  & $0.386\pm 0.062$ & $0.398\pm 0.060$ & $0.315\pm 0.060$ & $0.400\pm0.057$ \\
\hline
\end{tabular}
\end{ruledtabular}
\end{table*}
\begin{table*}
\caption
[tab:ftk0sn1]{
 Legendre polynomial coefficients for 
 the $K^- p \to  {\bar K}^0 n$ reaction for the four lowest beam momenta.
 } \label{tab:ftk0sn1}
\begin{ruledtabular}
\begin{tabular}{|c|c|c|c|c|} 
\hline
 $p_{K^-}\pm \delta_p$ [MeV/$c$] & $514\pm 10$ & $560\pm 11$ &
                                     $581\pm 12$ & $629\pm 11$ \\
\hline
 $A_0$ & $0.3071\pm 0.0092$ & $0.3370\pm 0.0074$ & $0.3242\pm 0.0070$ & $0.3100\pm 0.0056$ \\
 $A_1$ & $-0.023\pm 0.021$ & $0.032\pm 0.017$ & $0.028\pm 0.016$ & $0.068\pm 0.014$ \\
 $A_2$ & $0.247\pm 0.026$ & $0.257\pm 0.023$ & $0.250\pm 0.021$ & $0.253\pm 0.018$ \\
 $A_3$ & $-0.188\pm 0.031$ & $-0.173\pm 0.027$ & $-0.160\pm 0.025$ & $-0.120\pm 0.021$ \\
 $A_4$ & $0.008\pm 0.031$ & $-0.004\pm 0.025$ & $0.095\pm 0.023$ & $0.092\pm 0.019$ \\
 $A_5$ & $0.002\pm 0.034$ & $0.021\pm 0.026$ & $0.039\pm 0.025$ & $0.036\pm 0.019$ \\
 $\chi^2$/ndf & 0.99 & 1.43 & 1.06 & 1.02 \\
\hline
\end{tabular}
\end{ruledtabular}
\end{table*}
\begin{table*}
\caption
[tab:ftk0sn2]{
 Legendre polynomial coefficients for 
 the $K^- p \to  {\bar K}^0 n$ reaction for the four highest beam momenta.
 } \label{tab:ftk0sn2}
\begin{ruledtabular}
\begin{tabular}{|c|c|c|c|c|} 
\hline
 $p_{K^-}\pm \delta_p$ [MeV/$c$] & $659\pm 12$ & $687\pm 11$ &
                                     $714\pm 11$ & $750\pm 13$ \\
\hline
 $A_0$ & $0.2774\pm 0.0055$ & $0.2643\pm 0.0054$ & $0.2241\pm 0.0049$ & $0.1912\pm 0.0042$ \\
 $A_1$ & $0.023\pm 0.013$ & $0.008\pm 0.013$ & $-0.057\pm 0.012$ & $0.021\pm 0.011$ \\
 $A_2$ & $0.208\pm 0.018$ & $0.207\pm 0.018$ & $0.194\pm 0.016$ & $0.170\pm 0.014$ \\
 $A_3$ & $-0.180\pm 0.020$ & $-0.161\pm 0.020$ & $-0.160\pm 0.019$ & $-0.111\pm 0.016$ \\
 $A_4$ & $0.072\pm 0.018$ & $0.088\pm 0.018$ & $0.090\pm 0.016$ & $0.086\pm 0.013$ \\
 $A_5$ & $0.0003\pm 0.0179$ & $0.018\pm 0.018$ & $-0.025\pm 0.016$ & $0.019\pm 0.012$ \\
 $\chi^2$/ndf & 0.50 & 1.48 & 0.63 & 1.92 \\
\hline
\end{tabular}
\end{ruledtabular}
\end{table*}
 The differential cross sections are given as a function of $\cos(\theta^*)$,
 where the full range from -1 to 1 is divided in 16 bins.
 The bins with the acceptance below 0.5\% are not
 presented.
 To calculate our cross sections,
 the PDG~\cite{PDG} branching ratios $0.358\pm 0.005$
 for the $\Lambda \to \pi^0 n$ decay and $0.3069\pm 0.0005$
 for the $K^0_S \to \pi^0\pi^0$ decay were used.
 The effective proton density in the target times
 the effective target length was
 $(4.05\pm0.08)\times 10^{-4}$~ mb$^{-1}$. 
 The calculation of the total number
 of beam kaons incident on the target
 is given in detail in Ref.~\cite{NakPhD}.
 The uncertainties in our results for the differential
 cross sections are statistical only. 
 The overall systematic uncertainty in the differential
 cross sections was estimated to be about 7\%;
 it is not included in the errors presented. 
 The major contributions to the systematic uncertainty
 come from the evaluation of the losses of good
 events due to pile-up clusters and from the uncertainty in
 the total number of beam kaons incident on the target. 
\begin{figure*}
\includegraphics[width=16.cm,height=15.cm,bbllx=0.5cm,bblly=1.cm,bburx=19.5cm,bbury=18.5cm]{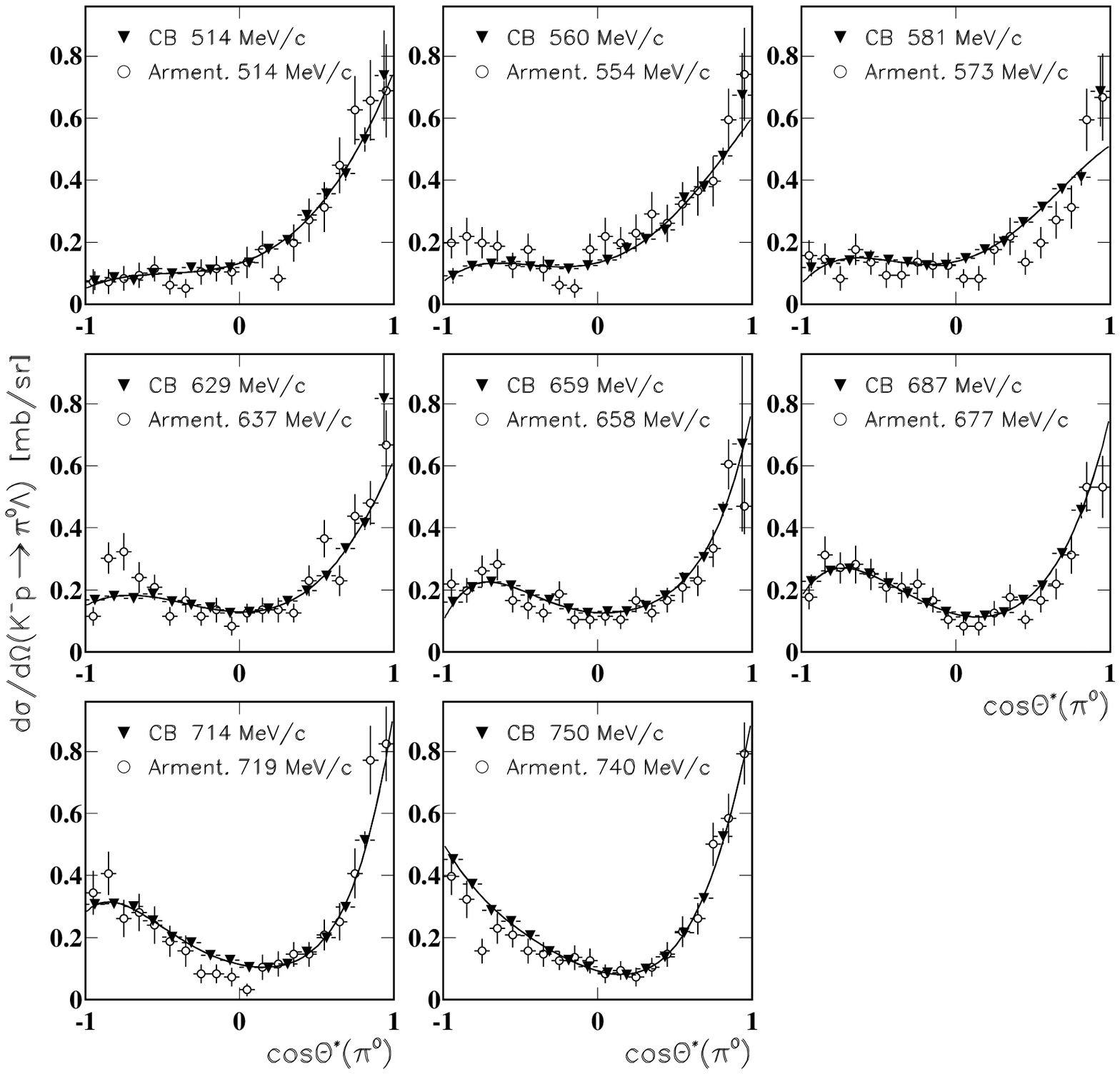}
\caption{
 Our differential cross sections for $K^-p \to \pi^0 \Lambda$ compared
 to the results of Armenteros~\protect\cite{Arm70} for similar beam momenta. 
 The curves are the Legendre polynomial fits of our data. 
}
 \label{fig:lampi0_dxs_v13_arm} 
\end{figure*}
\begin{table*}
\caption
[tab:lpi01]{
 Differential cross sections for the $K^- p \to \pi^0 \Lambda$ reaction
 for the four lowest beam momenta.
 } \label{tab:lpi01}
\begin{ruledtabular}
\begin{tabular}{|c|c|c|c|c|} 
\hline
 $p_{K^-}\pm \delta_p$ [MeV/$c$] & $514\pm 10$ & $560\pm 11$ &
                                     $581\pm 12$ & $629\pm 11$ \\
\hline
  $\cos\theta^{\ast}(\pi^0)$ & $d\sigma/d\Omega$ [mb/sr] &
  $d\sigma/d\Omega$ [mb/sr]&  $d\sigma/d\Omega$ [mb/sr] &  $d\sigma/d\Omega$ [mb/sr]\\
\hline
 -0.94 & $0.076\pm 0.031$ & $0.092\pm 0.024$ & $0.118\pm 0.028$ & $0.167\pm0.016$ \\
 -0.81 & $0.086\pm 0.012$ & $0.123\pm 0.010$ & $0.132\pm 0.011$ & $0.180\pm0.008$ \\
 -0.69 & $0.078\pm 0.011$ & $0.130\pm 0.009$ & $0.141\pm 0.008$ & $0.174\pm0.007$ \\
 -0.56 & $0.100\pm 0.010$ & $0.138\pm 0.008$ & $0.153\pm 0.008$ & $0.188\pm0.007$ \\
 -0.44 & $0.098\pm 0.009$ & $0.122\pm 0.008$ & $0.143\pm 0.008$ & $0.163\pm0.006$ \\
 -0.31 & $0.118\pm 0.010$ & $0.128\pm 0.008$ & $0.136\pm 0.007$ & $0.151\pm0.006$ \\
 -0.19 & $0.112\pm 0.010$ & $0.115\pm 0.008$ & $0.126\pm 0.007$ & $0.144\pm0.006$ \\
 -0.06 & $0.119\pm 0.010$ & $0.125\pm 0.008$ & $0.128\pm 0.007$ & $0.126\pm0.006$ \\
 0.06 & $0.134\pm 0.012$ & $0.143\pm 0.008$ & $0.149\pm 0.008$ & $0.129\pm0.006$ \\
 0.19 & $0.178\pm 0.013$ & $0.179\pm 0.009$ & $0.176\pm 0.008$ & $0.140\pm0.006$ \\
 0.31 & $0.206\pm 0.014$ & $0.211\pm 0.010$ & $0.202\pm 0.009$ & $0.164\pm0.006$ \\
 0.44 & $0.288\pm 0.017$ & $0.241\pm 0.011$ & $0.265\pm 0.010$ & $0.199\pm0.007$ \\
 0.56 & $0.357\pm 0.018$ & $0.345\pm 0.014$ & $0.315\pm 0.012$ & $0.245\pm0.009$ \\
 0.69 & $0.421\pm 0.023$ & $0.379\pm 0.018$ & $0.373\pm 0.014$ & $0.334\pm0.011$ \\
 0.81 & $0.531\pm 0.039$ & $0.477\pm 0.028$ & $0.410\pm 0.027$ & $0.414\pm0.020$ \\
 0.94 & $0.737\pm 0.145$ & $0.675\pm 0.135$ & $0.686\pm 0.113$ & $0.817\pm0.150$ \\
\hline
\end{tabular}
\end{ruledtabular}
\end{table*}
\begin{table*}
\caption
[tab:lpi02]{
 Differential cross sections for the $K^- p \to \pi^0 \Lambda$ reaction
 for the four highest beam momenta.
 } \label{tab:lpi02}
\begin{ruledtabular}
\begin{tabular}{|c|c|c|c|c|} 
\hline
 $p_{K^-}\pm \delta_p$ [MeV/$c$] & $659\pm 12$ & $687\pm 11$ &
                                     $714\pm 11$ & $750\pm 13$ \\
\hline
  $\cos\theta^{\ast}(\pi^0)$ & $d\sigma/d\Omega$ [mb/sr] &
  $d\sigma/d\Omega$ [mb/sr]&  $d\sigma/d\Omega$ [mb/sr] &  $d\sigma/d\Omega$ [mb/sr]\\
\hline
 -0.94 & $0.162\pm 0.016$ & $0.228\pm 0.021$ & $0.307\pm 0.017$ & $0.452\pm0.016$ \\
 -0.81 & $0.209\pm 0.009$ & $0.262\pm 0.010$ & $0.308\pm 0.009$ & $0.372\pm0.008$ \\
 -0.69 & $0.226\pm 0.008$ & $0.268\pm 0.009$ & $0.300\pm 0.008$ & $0.288\pm0.007$ \\
 -0.56 & $0.214\pm 0.008$ & $0.252\pm 0.008$ & $0.254\pm 0.007$ & $0.253\pm0.006$ \\ 
 -0.44 & $0.183\pm 0.006$ & $0.221\pm 0.007$ & $0.202\pm 0.007$ & $0.206\pm0.005$ \\ 
 -0.31 & $0.168\pm 0.006$ & $0.189\pm 0.007$ & $0.185\pm 0.006$ & $0.156\pm0.005$ \\ 
 -0.19 & $0.140\pm 0.006$ & $0.157\pm 0.006$ & $0.142\pm 0.005$ & $0.127\pm0.004$ \\ 
 -0.06 & $0.125\pm 0.005$ & $0.130\pm 0.005$ & $0.127\pm 0.005$ & $0.105\pm0.004$ \\ 
 0.06 & $0.131\pm 0.005$ & $0.114\pm 0.005$ & $0.104\pm 0.005$ & $0.087\pm0.004$ \\ 
 0.19 & $0.131\pm 0.005$ & $0.116\pm 0.005$ & $0.102\pm 0.005$ & $0.079\pm0.003$ \\ 
 0.31 & $0.149\pm 0.006$ & $0.128\pm 0.006$ & $0.115\pm 0.005$ & $0.100\pm0.004$ \\ 
 0.44 & $0.182\pm 0.007$ & $0.168\pm 0.007$ & $0.154\pm 0.006$ & $0.138\pm0.005$ \\ 
 0.56 & $0.239\pm 0.008$ & $0.214\pm 0.008$ & $0.200\pm 0.007$ & $0.216\pm0.006$ \\ 
 0.69 & $0.306\pm 0.011$ & $0.317\pm 0.011$ & $0.298\pm 0.011$ & $0.327\pm0.009$ \\ 
 0.81 & $0.461\pm 0.023$ & $0.457\pm 0.026$ & $0.514\pm 0.028$ & $0.526\pm0.026$ \\ 
 0.94 & $0.67\pm 0.28$ & --- & --- & --- \\  
\hline
\end{tabular}
\end{ruledtabular}
\end{table*}
\begin{table*}
\caption
[tab:ftlpi01]{
 Legendre polynomial coefficients for 
 the $K^- p \to \pi^0 \Lambda$ reaction for the four lowest beam momenta.
 } \label{tab:ftlpi01}
\begin{ruledtabular}
\begin{tabular}{|c|c|c|c|c|} 
\hline
 $p_{K^-}\pm \delta_p$ [MeV/$c$] & $514\pm 10$ & $560\pm 11$ &
                                     $581\pm 12$ & $629\pm 11$ \\
\hline
 $A_0$ & $0.2240\pm 0.0065$ & $0.2195\pm 0.0050$ & $0.2160\pm 0.0046$ & $0.2171\pm 0.0036$ \\
 $A_1$ & $0.274\pm 0.015$ & $0.208\pm 0.011$ & $0.178\pm 0.010$ & $0.1367\pm 0.0084$ \\
 $A_2$ & $0.180\pm 0.022$ & $0.148\pm 0.017$ & $0.120\pm 0.016$ & $0.172\pm 0.012$ \\
 $A_3$ & $0.075\pm 0.019$ & $0.056\pm 0.014$ & $0.044\pm 0.013$ & $0.098\pm 0.010$ \\
 $A_4$ & $-0.004\pm 0.019$ & $-0.031\pm 0.015$ & $-0.047\pm 0.014$ & $-0.052\pm 0.011$ \\
 $\chi^2$/ndf & 0.65 & 0.94 & 0.75 & 0.93 \\
\hline
\end{tabular}
\end{ruledtabular}
\end{table*}
\begin{table*}
\caption
[tab:ftlpi02]{
 Legendre polynomial coefficients for 
 the $K^- p \to \pi^0 \Lambda$ reaction for the four highest beam momenta.
 } \label{tab:ftlpi02}
\begin{ruledtabular}
\begin{tabular}{|c|c|c|c|c|} 
\hline
 $p_{K^-}\pm \delta_p$ [MeV/$c$] & $659\pm 12$ & $687\pm 11$ &
                                     $714\pm 11$ & $750\pm 13$ \\
\hline
 $A_0$ & $0.2291\pm 0.0045$ & $0.2411\pm 0.0049$ & $0.2527\pm 0.0051$ & $0.2619\pm 0.0044$ \\
 $A_1$ & $0.138\pm 0.012$ & $0.098\pm 0.014$ & $0.094\pm 0.014$ & $0.067\pm 0.013$ \\
 $A_2$ & $0.208\pm 0.016$ & $0.235\pm 0.018$ & $0.310\pm 0.018$ & $0.383\pm 0.016$ \\
 $A_3$ & $0.161\pm 0.018$ & $0.172\pm 0.021$ & $0.188\pm 0.020$ & $0.135\pm 0.018$ \\
 $A_4$ & $0.004\pm 0.014$ & $-0.004\pm 0.014$ & $0.041\pm 0.014$ & $0.060\pm 0.012$ \\
 $A_5$ & $0.041\pm 0.013$ & $0.021\pm 0.014$ & $0.042\pm 0.013$ & $-0.0003\pm 0.0111$ \\
 $\chi^2$/ndf & 0.52 & 0.29 & 1.60 & 1.15 \\
\hline
\end{tabular}
\end{ruledtabular}
\end{table*}

 Our differential cross sections for $K^- p \to {\bar K}^0 n$
 as a function of $\cos(\theta^*)$ for ${\bar K}^0$
 are given for each of the eight beam momenta in
 Tables~\ref{tab:k0sn1} and \ref{tab:k0sn2}. 
 Our differential cross sections for $K^-p \to \bar{K}^0 n$ are compared
 in Fig.~\ref{fig:k0sn_dxs_v13_arm_als} to the results of
 Armenteros~\cite{Arm70} and Alston-Garnjost~\cite{Alston} for similar
 beam momenta. The results of Armenteros~\cite{Arm70} are corrected for
 the currently accepted branching ratio of 0.692
 for the $K^0_S \to \pi^+\pi^-$ decay mode~\cite{PDG}.
 For the major part of the spectra, our data are
 in agreement with the existing measurements within the error bars.
 Some small disagreements among the three data sets could be also
 because of the difference in beam momentum. 
 The curves in the figure are the Legendre
 polynomial fits of our $K^- p \to {\bar K}^0 n$ data,
\begin{equation}
 d\sigma/d\Omega=\sum_{l=0}^{l_{\mathrm{max}}}A_lP_l(\cos\theta^*),
\label{eqn:legpol}
\end{equation}
 where $P_l$ is the Legendre polynomial of order $l$, and $A_l$ is
 its coefficient. The maximum order $l_{\mathrm{max}}$ was chosen to
 provide a good fit to the data for each of the eight momenta.
 This choice results sometimes, especially for the lower
 beam momenta, in the higher-order coefficients being
 consistent with zero.
 The results of Legendre polynomial fits for $K^- p \to {\bar K}^0 n$ are given
 for each of the eight beam momenta in
 Tables~\ref{tab:ftk0sn1} and \ref{tab:ftk0sn2}.    
\begin{figure*}
\includegraphics[width=16.cm,height=15.cm,bbllx=0.5cm,bblly=1.cm,bburx=19.5cm,bbury=18.5cm]{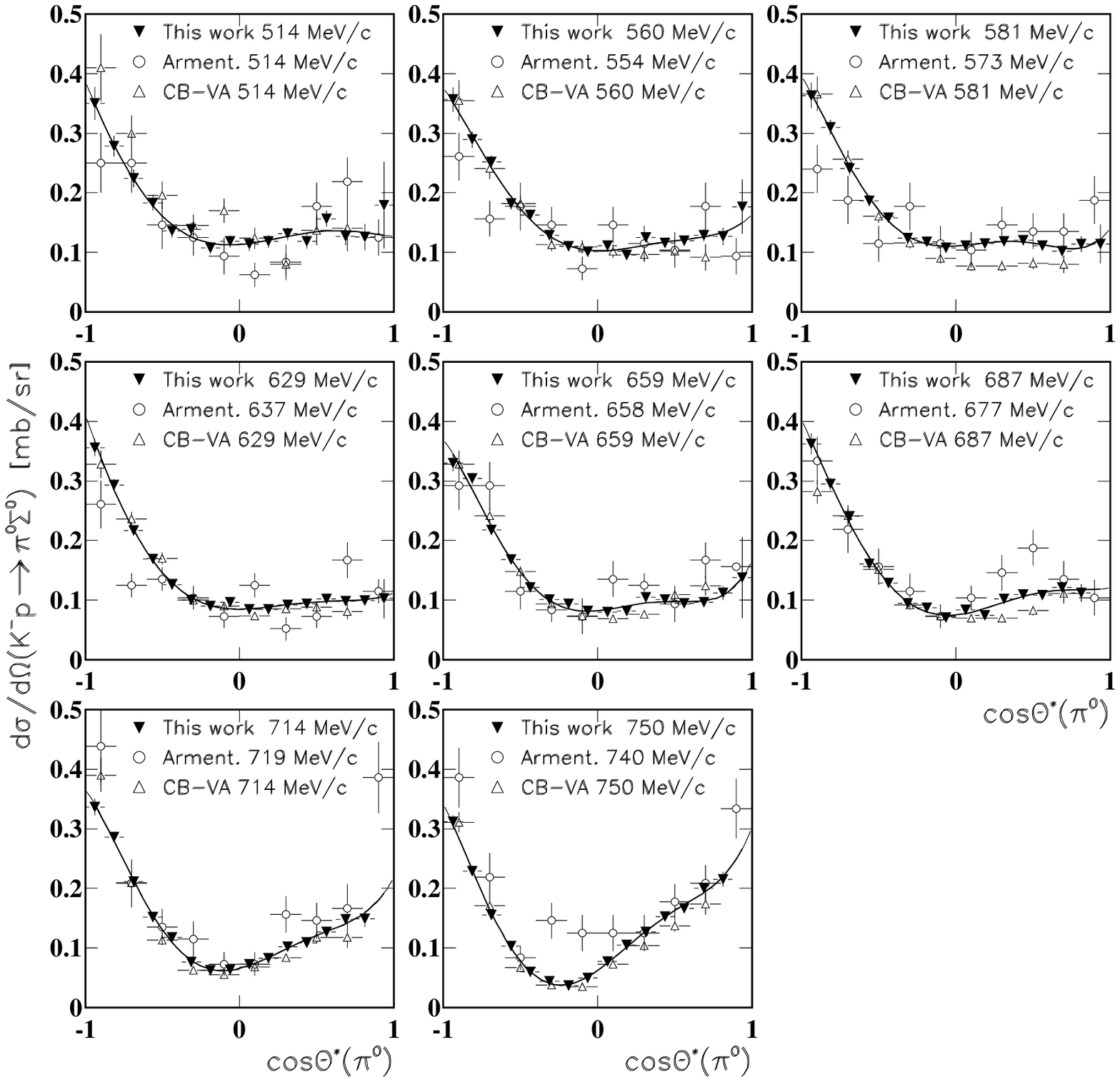}
\caption{
 Our differential cross sections for $K^-p \to \pi^0 \Sigma^0$ compared
 to the results of Armenteros~\protect\cite{Arm70} for similar beam momenta
 and to the VA results~\protect\cite{SPi0}. 
 The curves are the Legendre polynomial fits of our data. 
}
 \label{fig:sigpi0_dxs_v13_arm} 
\end{figure*}
\begin{table*}
\caption
[tab:spi01]{
 Differential cross sections for the $K^- p \to \pi^0 \Sigma^0$ reaction
 for the four lowest beam momenta.
 } \label{tab:spi01}
\begin{ruledtabular}
\begin{tabular}{|c|c|c|c|c|} 
\hline
 $p_{K^-}\pm \delta_p$ [MeV/$c$] & $514\pm 10$ & $560\pm 11$ &
                                     $581\pm 12$ & $629\pm 11$ \\
\hline
  $\cos\theta^{\ast}(\pi^0)$ & $d\sigma/d\Omega$ [mb/sr] &
  $d\sigma/d\Omega$ [mb/sr]&  $d\sigma/d\Omega$ [mb/sr] &  $d\sigma/d\Omega$ [mb/sr]\\
\hline
 -0.94 & $0.350\pm 0.028$ & $0.356\pm 0.021$ & $0.363\pm 0.021$ & $0.356\pm0.014$ \\ 
 -0.81 & $0.278\pm 0.017$ & $0.289\pm 0.013$ & $0.310\pm 0.012$ & $0.294\pm0.010$ \\
 -0.69 & $0.224\pm 0.015$ & $0.252\pm 0.011$ & $0.241\pm 0.011$ & $0.217\pm0.008$ \\
 -0.56 & $0.183\pm 0.013$ & $0.182\pm 0.010$ & $0.187\pm 0.010$ & $0.169\pm0.007$ \\
 -0.44 & $0.136\pm 0.011$ & $0.162\pm 0.009$ & $0.159\pm 0.008$ & $0.126\pm0.006$ \\
 -0.31 & $0.139\pm 0.011$ & $0.128\pm 0.008$ & $0.124\pm 0.008$ & $0.100\pm0.006$ \\
 -0.19 & $0.108\pm 0.010$ & $0.110\pm 0.008$ & $0.118\pm 0.007$ & $0.091\pm0.005$ \\
 -0.06 & $0.119\pm 0.010$ & $0.101\pm 0.007$ & $0.108\pm 0.007$ & $0.096\pm0.005$ \\
 0.06 & $0.114\pm 0.010$ & $0.111\pm 0.007$ & $0.111\pm 0.006$ & $0.085\pm0.005$ \\
 0.19 & $0.118\pm 0.010$ & $0.096\pm 0.007$ & $0.115\pm 0.006$ & $0.084\pm0.005$ \\
 0.31 & $0.131\pm 0.011$ & $0.125\pm 0.008$ & $0.119\pm 0.007$ & $0.092\pm0.005$ \\
 0.44 & $0.119\pm 0.012$ & $0.116\pm 0.008$ & $0.121\pm 0.006$ & $0.094\pm0.005$ \\
 0.56 & $0.157\pm 0.013$ & $0.120\pm 0.008$ & $0.111\pm 0.008$ & $0.103\pm0.006$ \\
 0.69 & $0.127\pm 0.014$ & $0.129\pm 0.010$ & $0.102\pm 0.010$ & $0.099\pm0.007$ \\
 0.81 & $0.126\pm 0.022$ & $0.127\pm 0.014$ & $0.114\pm 0.013$ & $0.099\pm0.010$ \\
 0.94 & $0.179\pm 0.073$ & $0.177\pm 0.045$ & $0.114\pm 0.033$ & $0.103\pm0.033$ \\
\hline
\end{tabular}
\end{ruledtabular}
\end{table*}
\begin{table*}
\caption
[tab:spi02]{
 Differential cross sections for the $K^- p \to \pi^0 \Sigma^0$ reaction
 for the four highest beam momenta.
 } \label{tab:spi02}
\begin{ruledtabular}
\begin{tabular}{|c|c|c|c|c|} 
\hline
 $p_{K^-}\pm \delta_p$ [MeV/$c$] & $659\pm 12$ & $687\pm 11$ &
                                     $714\pm 11$ & $750\pm 13$ \\
\hline
  $\cos\theta^{\ast}(\pi^0)$ & $d\sigma/d\Omega$ [mb/sr] &
  $d\sigma/d\Omega$ [mb/sr]&  $d\sigma/d\Omega$ [mb/sr] &  $d\sigma/d\Omega$ [mb/sr]\\
\hline
 -0.94 & $0.330\pm 0.014$ & $0.362\pm 0.017$ & $0.337\pm 0.014$ & $0.312\pm0.011$ \\
 -0.81 & $0.304\pm 0.009$ & $0.295\pm 0.010$ & $0.286\pm 0.009$ & $0.229\pm0.006$ \\
 -0.69 & $0.218\pm 0.008$ & $0.241\pm 0.008$ & $0.211\pm 0.007$ & $0.155\pm0.005$ \\
 -0.56 & $0.168\pm 0.007$ & $0.161\pm 0.007$ & $0.152\pm 0.006$ & $0.103\pm0.004$ \\
 -0.44 & $0.121\pm 0.006$ & $0.129\pm 0.005$ & $0.118\pm 0.005$ & $0.060\pm0.003$ \\
 -0.31 & $0.101\pm 0.005$ & $0.095\pm 0.005$ & $0.076\pm 0.004$ & $0.044\pm0.003$ \\
 -0.19 & $0.094\pm 0.005$ & $0.087\pm 0.005$ & $0.062\pm 0.004$ & $0.037\pm0.003$ \\
 -0.06 & $0.082\pm 0.005$ & $0.070\pm 0.004$ & $0.064\pm 0.004$ & $0.049\pm0.003$ \\
 0.06 & $0.080\pm 0.005$ & $0.083\pm 0.005$ & $0.073\pm 0.004$ & $0.077\pm0.003$ \\
 0.19 & $0.082\pm 0.005$ & $0.075\pm 0.005$ & $0.083\pm 0.004$ & $0.105\pm0.004$ \\
 0.31 & $0.105\pm 0.005$ & $0.102\pm 0.006$ & $0.102\pm 0.005$ & $0.127\pm0.004$ \\
 0.44 & $0.101\pm 0.005$ & $0.109\pm 0.006$ & $0.109\pm 0.005$ & $0.153\pm0.005$ \\
 0.56 & $0.095\pm 0.005$ & $0.108\pm 0.006$ & $0.127\pm 0.006$ & $0.166\pm0.005$ \\
 0.69 & $0.097\pm 0.007$ & $0.122\pm 0.007$ & $0.148\pm 0.007$ & $0.200\pm0.007$ \\
 0.81 & $0.112\pm 0.012$ & $0.112\pm 0.015$ & $0.149\pm 0.013$ & $0.215\pm0.013$ \\
 0.94 & $0.138\pm 0.068$ & --- & --- & --- \\
\hline
\end{tabular}
\end{ruledtabular}
\end{table*}
\begin{table*}
\caption
[tab:ftspi01]{
 Legendre polynomial coefficients for 
 the $K^- p \to \pi^0 \Sigma^0$ reaction for the four lowest beam momenta.
 } \label{tab:ftspi01}
\begin{ruledtabular}
\begin{tabular}{|c|c|c|c|c|} 
\hline
 $p_{K^-}\pm \delta_p$ [MeV/$c$] & $514\pm 10$ & $560\pm 11$ &
                                     $581\pm 12$ & $629\pm 11$ \\
\hline
 $A_0$ & $0.1603\pm 0.0049$ & $0.1598\pm 0.0034$ & $0.1582\pm 0.0030$ & $0.1387\pm 0.0023$ \\
 $A_1$ & $-0.079\pm 0.013$ & $-0.0874\pm 0.0086$ & $-0.1026\pm 0.0076$ & $-0.1009\pm 0.0060$ \\
 $A_2$ & $0.095\pm 0.017$ & $0.112\pm 0.011$ & $0.1010\pm 0.0097$ & $0.1141\pm 0.0077$ \\
 $A_3$ & $-0.058\pm 0.020$ & $-0.035\pm 0.014$ & $-0.049\pm 0.012$ & $-0.0571\pm 0.0095$ \\
 $A_4$ & $0.002\pm 0.017$ & $-0.002\pm 0.011$ & $0.009\pm 0.010$ & $0.0089\pm 0.0079$ \\
 $A_5$ & $0.006\pm 0.018$ & $0.016\pm 0.012$ & $0.024\pm 0.011$ & $0.0083\pm 0.0088$ \\
 $\chi^2$/ndf & 0.85 & 1.03 & 0.27 & 0.82 \\
\hline
\end{tabular}
\end{ruledtabular}
\end{table*}
\begin{table*}
\caption
[tab:ftspi02]{
 Legendre polynomial coefficients for 
 the $K^- p \to \pi^0 \Sigma^0$ reaction for the four highest beam momenta.
 } \label{tab:ftspi02}
\begin{ruledtabular}
\begin{tabular}{|c|c|c|c|c|} 
\hline
 $p_{K^-}\pm \delta_p$ [MeV/$c$] & $659\pm 12$ & $687\pm 11$ &
                                     $714\pm 11$ & $750\pm 13$ \\
\hline
 $A_0$ & $0.1397\pm 0.0027$ & $0.1426\pm 0.0033$ & $0.1437\pm 0.0030$ & $0.1439\pm 0.0027$ \\
 $A_1$ & $-0.0897\pm 0.0073$ & $-0.0939\pm 0.0092$ & $-0.0547\pm 0.0083$ & $0.0185\pm 0.0074$ \\
 $A_2$ & $0.1202\pm 0.0095$ & $0.128\pm 0.012$ & $0.153\pm 0.011$ & $0.1693\pm 0.0095$ \\
 $A_3$ & $-0.041\pm 0.012$ & $-0.061\pm 0.014$ & $-0.046\pm 0.013$ & $-0.069\pm 0.011$ \\
 $A_4$ & $0.0076\pm 0.0089$ & $-0.009\pm 0.010$ & $-0.0044\pm 0.0094$ & $0.0085\pm 0.0081$ \\
 $A_5$ & $0.0286\pm 0.0095$ & $0.014\pm 0.011$ & $0.0270\pm 0.0095$ & $0.0313\pm 0.0078$ \\
 $\chi^2$/ndf & 1.79 & 1.95 & 0.89 & 0.96 \\
\hline
\end{tabular}
\end{ruledtabular}
\end{table*}

 Our differential cross sections for $K^- p \to \pi^0 \Lambda$
 as a function of $\cos(\theta^*)$ for $\pi^0$
 are given for each of the eight beam momenta in
 Tables~\ref{tab:lpi01} and \ref{tab:lpi02}.      
 Our differential cross sections for $K^-p \to \pi^0 \Lambda$ are compared
 in Fig.~\ref{fig:lampi0_dxs_v13_arm} to the results of
 Armenteros~\cite{Arm70} for similar beam momenta.
 The results of Armenteros are corrected for
 the currently accepted branching ratio of 0.639
 for the $\Lambda \to \pi^- p$ decay mode~\cite{PDG}.
 For the major part of the spectra, our data are
 in agreement with the results of Armenteros within the error bars.
 Note that our data have much smaller statistical uncertainties. 
 Similar to the $K^-p \to \bar{K}^0 n$ results,
 some disagreement between the data could come from the difference
 between our beam momenta and those of Armenteros.
 Also, there could be some $K^-p \to \pi^0 \Sigma^0$ background
 in the data from Armenteros, especially in the backward 
 $\theta^*$ angles where the background reaction has a larger yield. 
 The curves in the figure are the Legendre
 polynomial fits of our $K^-p \to \pi^0 \Lambda$ data.
 The fit results are given for each of the eight beam momenta in
 Tables~\ref{tab:ftlpi01} and \ref{tab:ftlpi02}.    

 Our differential cross sections for $K^- p \to \pi^0 \Sigma^0$
 as a function of $\cos(\theta^*)$ for $\pi^0$
 are given for each of the eight beam momenta in
 Tables~\ref{tab:spi01} and \ref{tab:spi02}.      
 Our differential cross sections for $K^-p \to \pi^0 \Sigma^0$ are compared
 in Fig.~\ref{fig:sigpi0_dxs_v13_arm} to the results of
 Armenteros~\cite{Arm70} for similar beam momenta and also
 to the results of an independent analysis of the same data set
 reported recently in Ref.~\cite{SPi0}.
 The results of Armenteros are corrected for
 the currently accepted branching ratio 0.639
 of the $\Lambda \to \pi^- p$ decay mode~\cite{PDG}.
 The curves in the figure are the Legendre
 polynomial fits of our $K^- p \to \pi^0 \Sigma^0$ data. 
 The fit results are given for each of the eight beam momenta in
 Tables~\ref{tab:ftspi01} and \ref{tab:ftspi02}.    
 Note that our data are in reasonable agreement with the data
 from Armenteros within the error bars,
 but the quality and statistics of our
 data are much better. The measurement
 of $K^-p \to \pi^0 \Sigma^0$ by Armenteros~\cite{Arm70}
 was conducted using a bubble-chamber
 set-up, in which the two photons from the $\pi^0$ decay and
 the photon from the $\Sigma^0$ decay were not detected.
 This resulted in a poorer $\theta^{\ast}$ resolution and in
 a significant background from the $\pi^0\pi^0\Lambda$ and
 $\pi^0\pi^0\Sigma^0$ final states, especially for the highest momenta
 where the yield of these background reactions becomes large.

 We note that some of our $K^-p \to \pi^0 \Sigma^0$ results,
 particularly for the forward production angles,
 are in disagreement with a separate analysis~\cite{SPi0} of
 the same data by the Valparaiso-Argonne (VA)
 group of the Crystal Ball collaboration at the AGS.
 The results presented in our paper are obtained by the UCLA group,
 several members of which disagreed with the VA analysis
 and are not in the authors of Ref.~\cite{SPi0}.
 There are several major differences between the two
 analyses that could lead to the disagreement observed
 in the $K^-p \to \pi^0 \Sigma^0$ results.
 First, we used the full fiducial volume of the CB
 in our analysis. This required a fine tuning of
 the MC simulation to reproduce the experimental conditions,
 including the CB energy trigger.
 The reduced fiducial volume, which is less sensitive
 to the MC simulation, was used in the VA analysis.
 However, this led to much less data and large uncertainties,
 especially for the forward direction where the largest discrepancy
 between the two analyses was observed.
 Second, the mean values of the beam-momentum spectra
 that were reconstructed by us for $K^-p$ interactions in
 the LH$_2$ target~\cite{l2pi0,3pi0lam} were used in the VA analysis
 as nominal beam momenta. The nominal momenta are used
 as the input for the transport matrix of the dipole magnet
 in order to calculate the momentum of an individual beam
 particle based on the information of the drift chambers
 that were located upstream and downstream of the magnet.
 Because of the energy losses in the beam counters,
 there is a difference between the beam momenta in the 
 magnet and in the target. The use of the incorrect
 beam-momentum information in the event reconstruction
 usually leads to some distortion
 of the final angular distributions in the c.m. system.
 Third, our analysis showed good agreement with the older
 measurements of $K^-p \to \bar{K}^0 n$ and $K^-p \to \pi^0 \Lambda$,
 the quality of which is much better than the old
 $K^- p \to \pi^0\Sigma^0$ data. 
 These measurements are also needed for the accurate
 subtraction of background contributions
 from the $K^- p \to \pi^0\Sigma^0$ candidates.
 The background subtraction in the VA analysis is
 less accurate as it relies on the Armenteros data~\cite{Arm70},
 which were obtained often for different beam momenta,
 with poor statistics, and without background subtraction.
 The same is for the subtraction of the $K^-p \to \pi^0 \pi^0 \Lambda$
 background; the poor Armenteros data~\cite{Arm70} were used
 in the VA analysis instead of the results obtained
 for this reaction in the same experiment~\cite{l2pi0}.
 Another difference between the two
 analyses is that we used the kinematic-fit technique
 as part of the event reconstruction.
 This provided a better experimental resolution in $\theta^{\ast}$
 and allows to use the fit C.L. for the event selection.

\begin{figure}
\includegraphics[width=8.cm,height=12.cm,bbllx=1.cm,bblly=0.5cm,bburx=14.5cm,bbury=21.5cm]{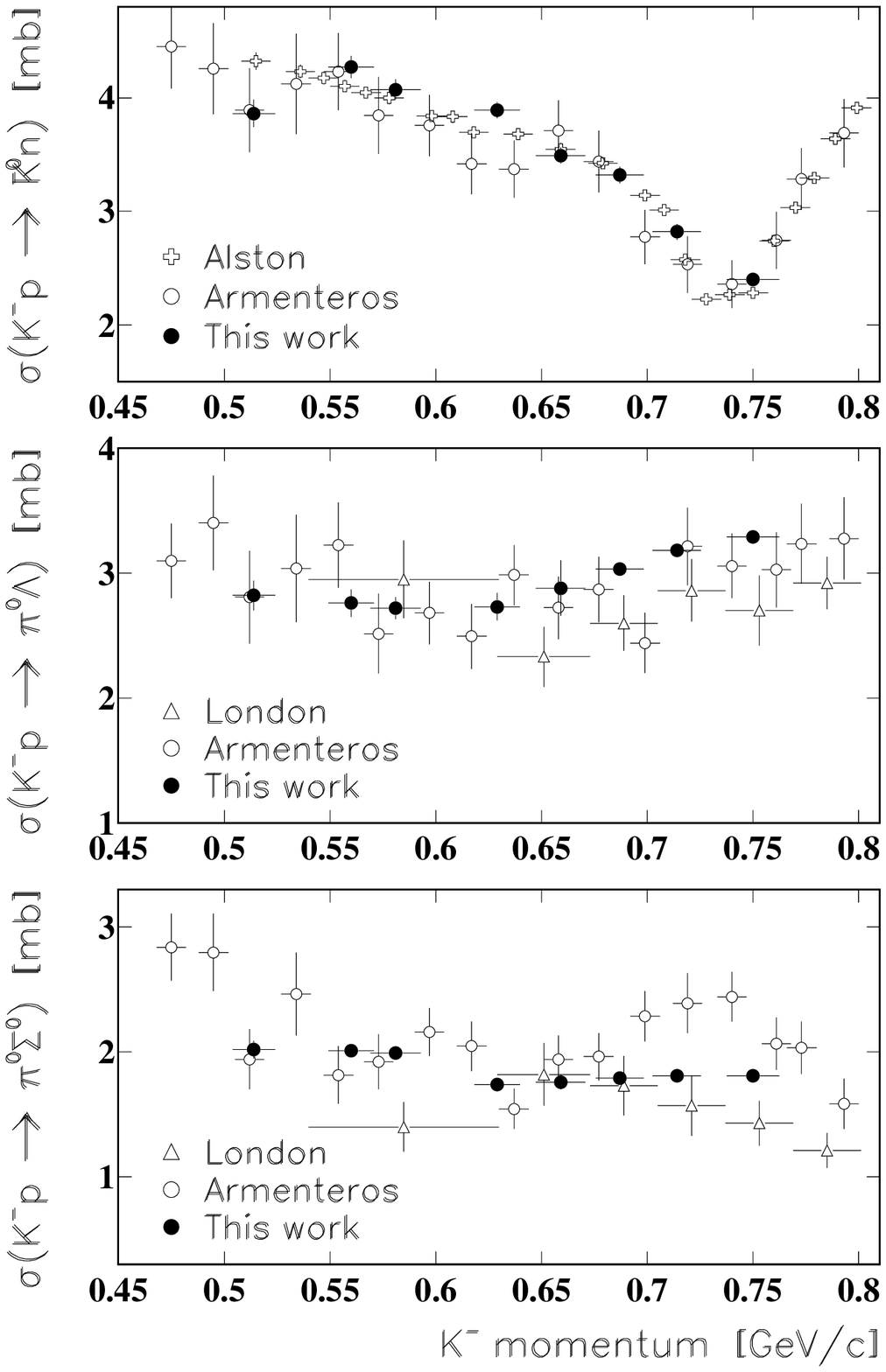}
\caption{
 Our total cross sections for $K^- p \to \bar{K}^0 n$, $K^-p \to \pi^0 \Lambda$,
 and $K^-p \to \pi^0 \Sigma^0$ compared to the results from
 Armenteros~\protect\cite{Arm70}, Alston-Garnjost~\protect\cite{Alston_tot},
 and London~\protect\cite{London75}.
}
 \label{fig:k0sn_lpi0_spi0_arm_als_xst_v17} 
\end{figure}

 Our evaluation of the total cross sections for reactions
 $K^- p \to {\bar K}^0 n$, $K^-p \to \pi^0 \Lambda$, and
 $K^- p \to \pi^0\Sigma^0$ was based on the Legendre
 polynomial fit of the corresponding differential cross
 sections; the results obtained are listed
 for each of the eight beam momenta in Table~\ref{tab:totxsc}.
\begin{table*}
\caption
[tab:totxsc]{
 The total cross sections for reactions
 $K^- p \to {\bar K}^0 n$, $K^- p \to \pi^0 \Lambda$, and 
 $K^- p \to \pi^0 \Sigma^0$ at the eight beam momenta.
 } \label{tab:totxsc}
\begin{ruledtabular}
\begin{tabular}{|l|c|c|c|c|c|c|c|c|} 
\hline
 $p_{K^-}\pm \sigma_p$ &
 514$\pm$10 & 560$\pm$11 & 581$\pm$12 & 629$\pm$11 &
 659$\pm$12 & 687$\pm$11 & 714$\pm$11 & 750$\pm$13 \\

[MeV/$c$] &  &  &  &  &  &  &  &  \\   
\hline
 $\sigma_{{\bar K}^0 n}$ [mb] & $3.86\pm0.12$ & $4.23\pm0.09$ & $4.07\pm0.09$ 
  & $3.89\pm0.07$ & $3.49\pm0.07$ & $3.32\pm0.07$ & $2.82\pm0.07$ & $2.40\pm0.06$ \\
 $\sigma_{\pi^0 \Lambda}$ [mb] & $2.82\pm0.12$ & $2.76\pm0.11$ & $2.72\pm0.09$
  & $2.73\pm0.11$ & $2.88\pm0.22$ & $3.03\pm0.04$ & $3.18\pm0.04$ & $3.29\pm0.04$ \\
 $\sigma_{\pi^0 \Sigma^0}$ [mb] & $2.02\pm0.07$ & $2.01\pm0.05$ & $1.99\pm0.04$
  & $1.74\pm0.04$ & $1.76\pm0.06$ & $1.79\pm0.03$ & $1.81\pm0.03$ & $1.81\pm0.02$ \\
\hline
\end{tabular}
\end{ruledtabular}
\end{table*}
 Since the results of our fits vary slightly
 depending on the maximum polynomial order used,
 there could be a small difference between our
 values for the total cross sections and the results
 of any independent analysis of our differential cross sections.
 The uncertainties that we give to our total cross sections 
 are based only on the statistical uncertainties in the points
 of the differential cross sections themselves.
 This leads to a smaller uncertainty of the total cross sections
 obtained from the differential cross sections in which
 the point $\cos(\theta^*)=0.94$ is omitted. 
 The systematic uncertainty of 7\% in the differential cross
 sections is not included in the error calculation.
 Our values for the total cross sections of the three reactions
 are compared in Fig.~\ref{fig:k0sn_lpi0_spi0_arm_als_xst_v17} to the
 results from Armenteros~\cite{Arm70}, Alston-Garnjost~\cite{Alston_tot},
 and London~\cite{London75}.
 There is no comparison to
 the $K^-p \to \pi^0\Sigma^0$ total cross sections from
 the VA analysis as their values are not
 reported in Ref.~\cite{SPi0}.
 As shown, our total cross sections
 for $K^- p \to \bar{K}^0 n$ within the error bars are in
 agreement with the existing data.
 Our total cross sections for $K^-p \to \pi^0 \Lambda$
 are in a better agreement with the data from Armenteros
 than with the ones from
 London, where our results have the smallest statistical uncertainties.
 Our total cross sections for $K^-p \to \pi^0\Sigma^0$
 also have the smallest statistical uncertainties and
 lie somewhere between the results from Armenteros
 and London.
 
 The induced polarization of the $\Lambda$ hyperon in
 the $K^-p \to \pi^0 \Lambda$ reaction was measured via its
 decay asymmetry:
\begin{equation}
   P_{\Lambda}(\cos\theta^{\ast}) = 3(\sum_i \cos\xi_i)/(\alpha_\Lambda N(\theta^{\ast}))~,
\label{eqn:lampol}
\end{equation}
 where $\theta^{\ast}$  is the angle between the direction of the outgoing $\pi^0$
 and the incident $K^-$ in the c.m. system.
 The $\xi$ angle is defined by 
\begin{equation}
\label{eqn:xiang}
 \cos\xi = (\hat{\boldsymbol{K}}^- \times \hat{\boldsymbol{\pi}}^0)
 \cdot \hat{\boldsymbol{n}} /
 |\hat{\boldsymbol{K}}^- \times \hat{\boldsymbol{\pi}}^0|
 = \hat{\boldsymbol{N}} \cdot \hat{\boldsymbol{n}}~,
\end{equation}
 where $\hat{\boldsymbol{K}}^-$ and $\hat{\boldsymbol{\pi}}^0$
 are unit vectors in the direction of the incident $K^-$ and
 the outgoing $\pi^0$ meson respectively,
 $\hat{\boldsymbol{n}}$ is a unit vector in the direction of
 the decay neutron in the $\Lambda$ rest frame,
 $\hat{\boldsymbol{N}}$ is the normal to the production plane.
 $N(\theta^{\ast})$ is the total number of $\Lambda$ hyperons
 produced at the angle $\theta^{\ast}$,
 and $\alpha_\Lambda = +0.65$ is
 the asymmetry parameter for $\Lambda \to \pi^0 n$.

\begin{figure*}
\includegraphics[width=17.cm,height=9.cm,bbllx=1.cm,bblly=1.cm,bburx=19.5cm,bbury=10.5cm]{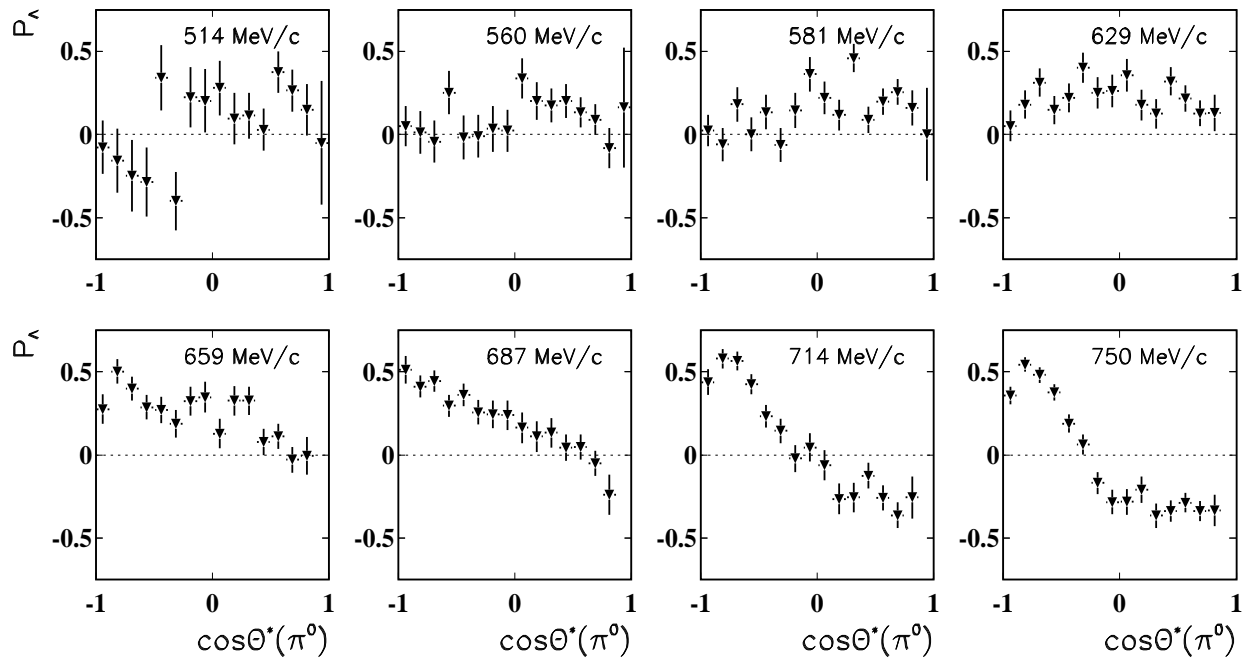}
\caption{
 Our results for the $\Lambda$ polarization as a function
 of $\cos\theta^{\ast}(\pi^0)$ for the $K^-p \to \pi^0 \Lambda$ reaction
  at the eight beam momenta.
}
 \label{fig:lampi0_pol_v17} 
\end{figure*}

\begin{figure*}
\includegraphics[width=15.5cm,height=14.cm,bbllx=0.5cm,bblly=1.cm,bburx=19.5cm,bbury=18.5cm]{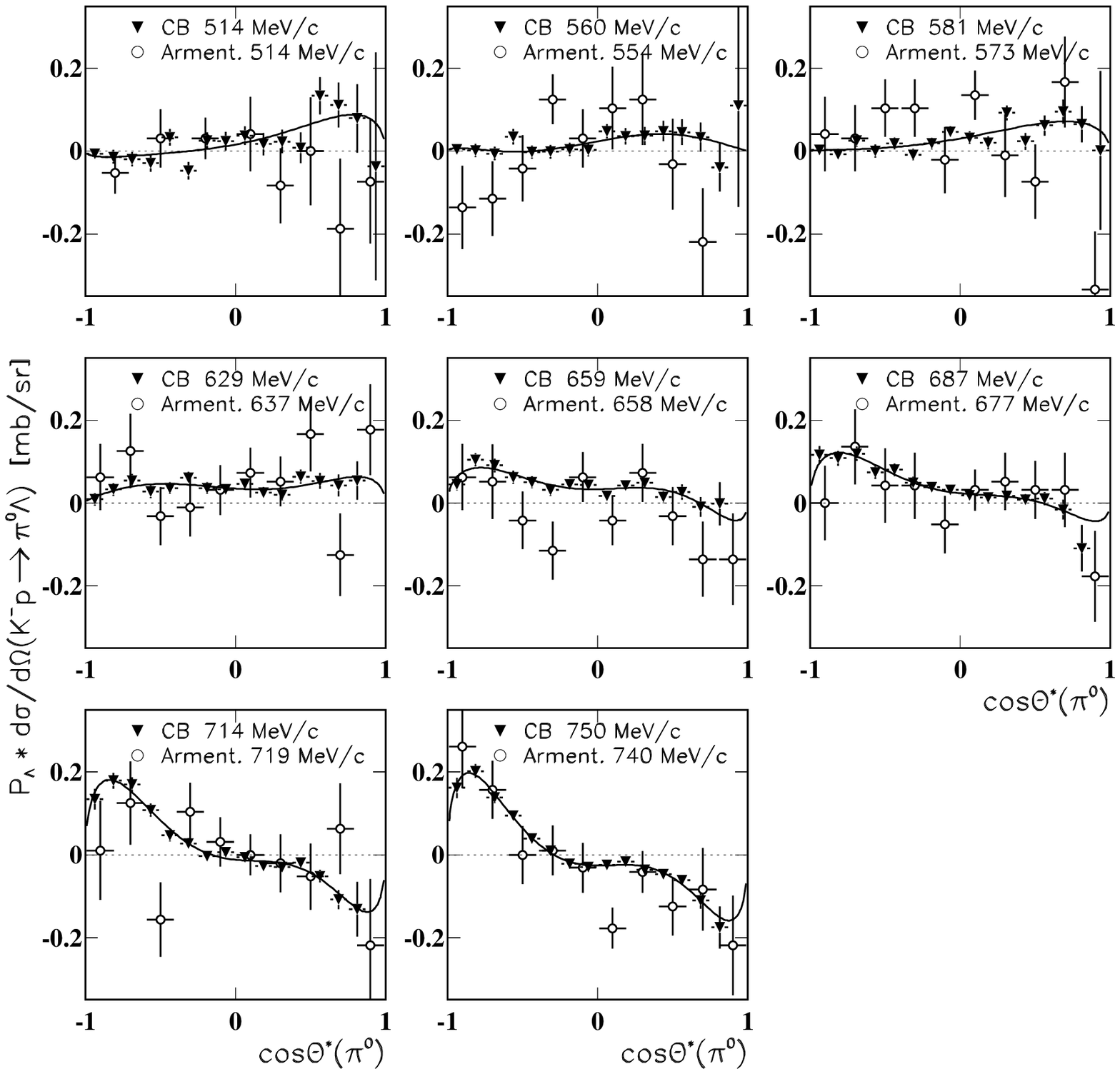}
\caption{
 Our results for the product of the $\Lambda$ polarization
 and the differential cross section of 
 $K^- p \to \pi^0 \Lambda$ compared to the data
 from Armenteros~\protect\cite{Arm70}.
 The curves are the fits of our data to
 the first associated Legendre functions. 
}
 \label{fig:lampi0_pol_v17_arm} 
\end{figure*}

 Our results for the $\Lambda$ polarization as a function
 of $\cos\theta^{\ast}(\pi^0)$ for the $K^-p \to \pi^0 \Lambda$ reaction
 are given for each of the eight beam momenta in
 Tables~\ref{tab:lpol1} and \ref{tab:lpol2}.
 These results are also shown in Fig.~\ref{fig:lampi0_pol_v17}.
\begin{table*}
\caption
[tab:lpol1]{
 Polarization of the $\Lambda$ hyperon in the $K^- p \to \pi^0 \Lambda$ reaction
 for the four lowest beam momenta.
 } \label{tab:lpol1}
\begin{ruledtabular}
\begin{tabular}{|c|c|c|c|c|} 
\hline
 $p_{K^-}\pm \delta_p$ [MeV/$c$] & $514\pm 10$ & $560\pm 11$ &
                                     $581\pm 12$ & $629\pm 11$ \\
\hline
  $\cos\theta^{\ast}(\pi^0)$ & $P_{\Lambda}$ &
  $P_{\Lambda}$ & $P_{\Lambda}$ & $P_{\Lambda}$ \\
\hline
 -0.94 & $-0.08\pm 0.16$ & $0.05\pm 0.12$ & $0.02\pm 0.09$ & $0.05\pm 0.09$ \\
 -0.81 & $-0.16\pm 0.19$ & $0.01\pm 0.13$ & $-0.06\pm 0.10$ & $0.18\pm 0.09$ \\
 -0.69 & $-0.25\pm 0.22$ & $-0.04\pm 0.13$ & $0.18\pm 0.10$ & $0.31\pm 0.09$ \\
 -0.56 & $-0.28\pm 0.21$ & $0.25\pm 0.13$ & $0.00\pm 0.10$ & $0.15\pm 0.08$ \\
 -0.44 & $0.34\pm 0.20$ &  $-0.02\pm 0.13$ & $0.13\pm 0.10$ & $0.22\pm 0.09$ \\
 -0.31 & $-0.40\pm 0.17$ & $-0.01\pm 0.13$ & $-0.06\pm 0.10$ & $0.40\pm 0.09$ \\
 -0.19 & $0.22\pm 0.18$ & $0.04\pm 0.14$ & $0.15\pm 0.11$ & $0.25\pm 0.09$ \\
 -0.06 & $0.20\pm 0.19$ & $0.02\pm 0.13$ & $0.36\pm 0.10$ & $0.26\pm 0.10$ \\
 0.06 & $0.28\pm 0.16$ & $0.34\pm 0.12$ & $0.22\pm 0.10$ & $0.36\pm 0.10$ \\
 0.19 & $0.10\pm 0.15$ & $0.20\pm 0.11$ & $0.12\pm 0.09$ & $0.18\pm 0.09$ \\ 
 0.31 & $0.11\pm 0.14$ & $0.18\pm 0.10$ & $0.46\pm 0.09$ & $0.12\pm 0.09$ \\
 0.44 & $0.03\pm 0.13$ & $0.20\pm 0.10$ & $0.09\pm 0.08$ & $0.32\pm 0.08$ \\
 0.56 & $0.38\pm 0.12$ & $0.13\pm 0.09$ & $0.20\pm 0.08$ & $0.22\pm 0.08$ \\ 
 0.69 & $0.26\pm 0.13$ & $0.09\pm 0.09$ & $0.26\pm 0.08$ & $0.13\pm 0.08$ \\
 0.81 & $0.15\pm 0.16$ & $-0.08\pm 0.12$ & $0.16\pm 0.11$ & $0.13\pm 0.11$ \\ 
 0.94 & $-0.05\pm 0.37$ & $0.16\pm 0.36$ & $0.00\pm 0.28$ & --- \\ 
\hline
\end{tabular}
\end{ruledtabular}
\end{table*}
\begin{table*}
\caption
[tab:lpol2]{
 Polarization of the $\Lambda$ hyperon in the $K^- p \to \pi^0 \Lambda$ reaction
 for the four highest beam momenta.
 } \label{tab:lpol2}
\begin{ruledtabular}
\begin{tabular}{|c|c|c|c|c|} 
\hline
 $p_{K^-}\pm \delta_p$ [MeV/$c$] & $659\pm 12$ & $687\pm 11$ &
                                     $714\pm 11$ & $750\pm 13$ \\
\hline
  $\cos\theta^{\ast}(\pi^0)$ & $P_{\Lambda}$ &
  $P_{\Lambda}$ & $P_{\Lambda}$ & $P_{\Lambda}$ \\
\hline
 -0.94 & $0.28\pm 0.09$ & $0.51\pm 0.08$ & $0.44\pm 0.08$ & $0.36\pm 0.05$ \\
 -0.81 & $0.50\pm 0.07$ & $0.41\pm 0.07$ & $0.58\pm 0.06$ & $0.54\pm 0.04$ \\
 -0.69 & $0.40\pm 0.07$ & $0.44\pm 0.06$ & $0.57\pm 0.06$ & $0.48\pm 0.05$ \\
 -0.56 & $0.29\pm 0.07$ & $0.30\pm 0.07$ & $0.43\pm 0.06$ & $0.38\pm 0.05$ \\
 -0.44 & $0.27\pm 0.08$ & $0.36\pm 0.07$ & $0.23\pm 0.07$ & $0.19\pm 0.06$ \\
 -0.31 & $0.19\pm 0.08$ & $0.26\pm 0.07$ & $0.15\pm 0.07$ & $0.06\pm 0.06$ \\
 -0.19 & $0.32\pm 0.09$ & $0.24\pm 0.08$ & $-0.02\pm 0.08$ & $-0.17\pm 0.07$ \\
 -0.06 & $0.35\pm 0.09$ & $0.24\pm 0.09$ & $0.05\pm 0.08$ & $-0.28\pm 0.07$ \\
 0.06 & $0.13\pm 0.09$ & $0.16\pm 0.09$ & $-0.06\pm 0.09$ & $-0.28\pm 0.08$ \\
 0.19 & $0.33\pm 0.09$ & $0.11\pm 0.09$ & $-0.26\pm 0.09$ & $-0.21\pm 0.08$ \\
 0.31 & $0.33\pm 0.09$ & $0.13\pm 0.09$ & $-0.26\pm 0.09$ & $-0.36\pm 0.07$ \\
 0.44 & $0.08\pm 0.08$ & $0.04\pm 0.08$ & $-0.13\pm 0.08$ & $-0.34\pm 0.06$ \\
 0.56 & $0.11\pm 0.08$ & $0.05\pm 0.08$ & $-0.26\pm 0.07$ & $-0.29\pm 0.06$ \\
 0.69 & $-0.03\pm 0.08$ & $-0.05\pm 0.08$ & $-0.36\pm 0.08$ & $-0.34\pm 0.06$ \\
 0.81 & $0.00\pm 0.11$ & $-0.24\pm 0.12$ & $-0.26\pm 0.13$ & $-0.33\pm 0.10$ \\
 0.94 & --- & --- & --- & --- \\ 
\hline
\end{tabular}
\end{ruledtabular}
\end{table*}
 The comparison of our results for
 the product of the $\Lambda$ polarization
 and the differential cross section of 
 $K^- p \to \pi^0 \Lambda$ with the data
 from Armenteros\cite{Arm70} is shown in
 Fig.~\ref{fig:lampi0_pol_v17_arm}.
 Both the results are in agreement within the error bars,
 where our statistical uncertainties are much smaller.
 The curves in the figure are the fits
 of our $K^- p \to \pi^0 \Lambda$ data
 to the first associated Legendre functions,
\begin{equation}
 P(d\sigma/d\Omega)=\sum_{l=0}^{l_{\mathrm{max}}}B_lP_l^1(\cos\theta^*).
\label{eqn:legfun}
\end{equation}
 The results of the fits are given
 for each of the eight beam momenta in
 Tables~\ref{tab:ftlpol1} and \ref{tab:ftlpol2}.    
\begin{table*}
\caption
[tab:ftlpol1]
{ Associated Legendre function coefficients for 
 the $K^- p \to \pi^0 \Lambda$ reaction
 for the four lowest beam momenta.
} \label{tab:ftlpol1}
\begin{ruledtabular}
\begin{tabular}{|c|c|c|c|c|} 
\hline
 $p_{K^-}\pm \delta_p$ [MeV/$c$] & $514\pm 10$ & $560\pm 11$ &
                                     $581\pm 12$ & $629\pm 11$ \\
\hline
 $B_1$ & $-0.0306\pm 0.0099$ & $-0.0232\pm 0.0072$ & $-0.0401\pm 0.0060$ & $-0.0489\pm 0.0054$ \\
 $B_2$ & $-0.0311\pm 0.0096$ & $-0.0114\pm 0.0067$ & $-0.0223\pm 0.0055$ & $-0.0051\pm 0.0051$ \\
 $B_3$ & $-0.0096\pm 0.0081$ & $0.0004\pm 0.0058$ & $-0.0062\pm 0.0047$ & $-0.0093\pm 0.0043$ \\
 $B_4$ & $-0.0034\pm 0.0051$ & $0.0044\pm 0.0040$ & $-0.0004\pm 0.0034$ & $-0.0052\pm 0.0035$ \\
 $\chi^2$/ndf & 1.40 & 0.69 & 1.61 & 1.30 \\
\hline
\end{tabular}
\end{ruledtabular}
\end{table*}
\begin{table*}
\caption
[tab:ftlpol2]{
 Associated Legendre function coefficients for 
 the $K^- p \to \pi^0 \Lambda$ reaction for the four highest beam momenta.
 } \label{tab:ftlpol2}
\begin{ruledtabular}
\begin{tabular}{|c|c|c|c|c|} 
\hline
 $p_{K^-}\pm \delta_p$ [MeV/$c$] & $659\pm 12$ & $687\pm 11$ &
                                     $714\pm 11$ & $750\pm 13$ \\
\hline
 $B_1$ & $-0.0435\pm 0.0054$ & $-0.0415\pm 0.0055$ & $-0.0111\pm 0.0053$ & $0.0038\pm 0.0043$ \\
 $B_2$ & $0.0249\pm 0.0058$ & $0.0407\pm 0.0060$ & $0.0750\pm 0.0059$ & $0.0785\pm 0.0048$ \\
 $B_3$ & $-0.0030\pm 0.0052$ & $-0.0099\pm 0.0055$ & $-0.0110\pm 0.0056$ & $-0.0124\pm 0.0048$ \\
 $B_4$ & $0.0109\pm 0.0043$ & $0.0112\pm 0.0043$ & $0.0251\pm 0.0042$ & $0.0312\pm 0.0033$ \\
 $B_5$ & $0.0030\pm 0.0033$ & $0.0017\pm 0.0035$ & $0.0035\pm 0.0036$ & $0.0019\pm 0.0029$ \\
 $\chi^2$/ndf & 1.18 & 0.61 & 0.97 & 0.57 \\
\hline
\end{tabular}
\end{ruledtabular}
\end{table*}

 The induced polarization of the $\Sigma^0$ hyperon in
 the $K^-p \to \pi^0 \Sigma^0$ reaction was measured via its
 decay asymmetry (see Ref.~\cite{Anders}):
\begin{equation}
\label{eqn:sigpol}
   P_{\Sigma^0}(\cos\theta^{\ast}) = -9(\sum_i \cos\xi_i \cos\psi_i)/(\alpha_\Lambda N(\theta^{\ast}))~,
\end{equation}
 where $\theta^{\ast}$  is the angle between the direction of the outgoing $\pi^0$
 and the incident $K^-$ in the c.m. system.
 The $\xi$ angle was defined by 
\begin{equation}
\label{eqn:xiang2}
 \cos~\xi = (\hat{\boldsymbol{K}}^- \times \hat{\boldsymbol{\Sigma}}^0)
 \cdot \hat{\boldsymbol{\Lambda}} /
 |\hat{\boldsymbol{K}}^- \times \hat{\boldsymbol{\Sigma}}^0|
 = \hat{\boldsymbol{N}} \cdot \hat{\boldsymbol{\Lambda}}~,
\end{equation}
 where $\hat{\boldsymbol{K}}^-$ and $\hat{\boldsymbol{\Sigma}}^0$
 are unit vectors in the direction of the incident $K^-$ and
 the outgoing $\Sigma^0$ hyperon, respectively.
 $\hat{\boldsymbol{\Lambda}}$ is a unit vector in the direction of
 the $\Lambda$ hyperon in the $\Sigma^0$ rest frame,
 $\hat{\boldsymbol{N}}$ is the normal to the production plane.
 The $\psi$ angle was defined by 
\begin{equation}
\label{eqn:psiang}
 \cos~\psi =
 \hat{\boldsymbol{\Lambda}} \cdot \hat{\boldsymbol{n}}~,
\end{equation}
 where $\hat{\boldsymbol{n}}$ is a unit vector in the direction of
 the decay neutron in the $\Lambda$ rest frame.
 $N(\theta^{\ast})$ is the total number of the $\Sigma^0$ hyperons
 produced at the angle $\theta^{\ast}$,
 and $\alpha_\Lambda = +0.65$ is
 the asymmetry parameter for $\Lambda \to \pi^0 n$.
 The measured values for the $\Sigma^0$ polarization were
 corrected for the $\cos~\xi$ acceptance, which turned out
 to be about 30\% smaller for the central $\cos~\xi$ values.
 (see Ref.~\cite{SPi0} for more details).

\begin{figure*}
\includegraphics[width=17.cm,height=9.cm,bbllx=1.cm,bblly=1.cm,bburx=19.5cm,bbury=10.5cm]{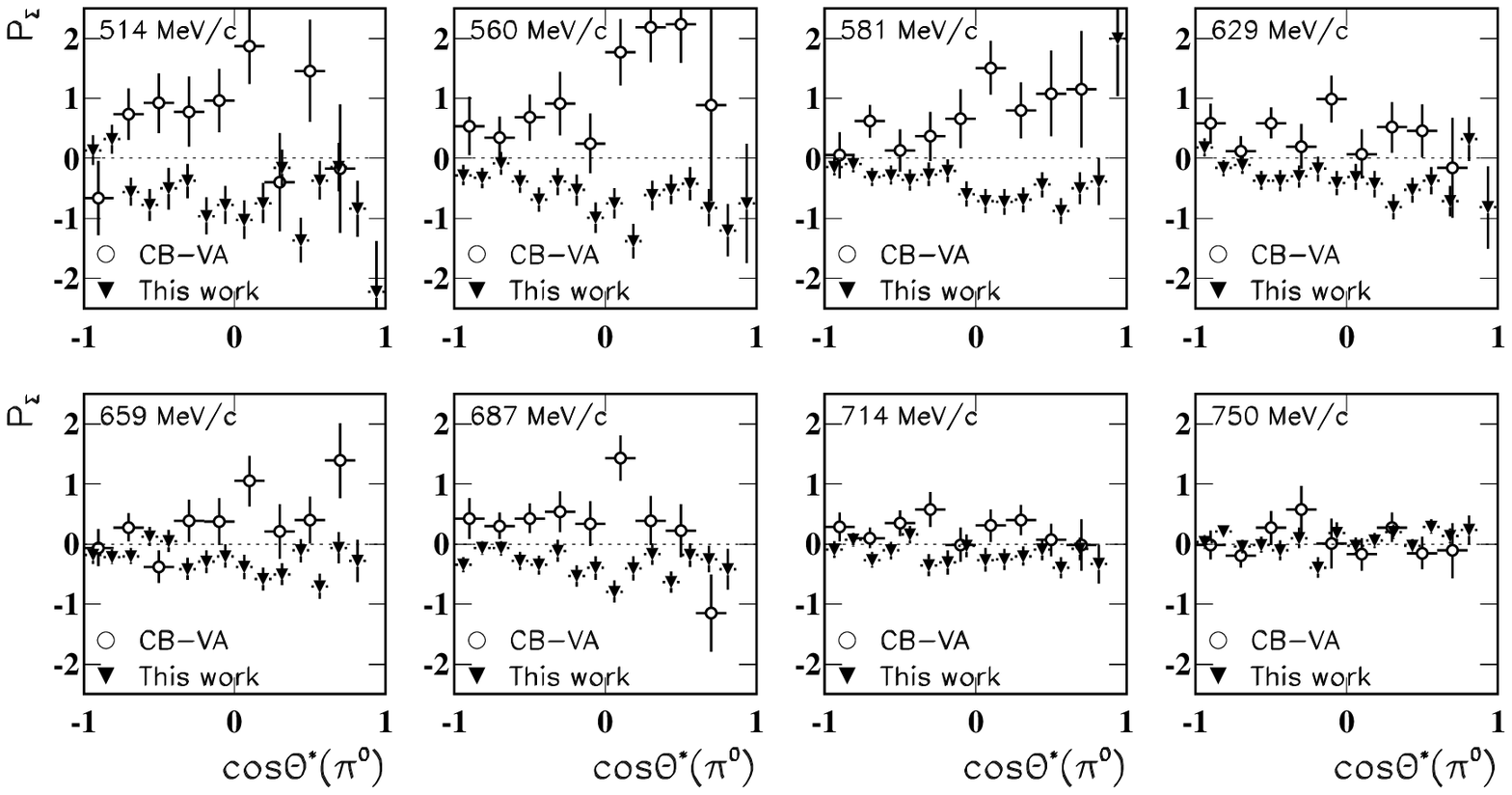}
\caption{
 Our results for the $\Sigma^0$ polarization as a function
 of $\cos\theta^{\ast}(\pi^0)$ for the $K^-p \to \pi^0 \Sigma^0$ reaction
 at the eight beam momenta, compared to the VA analysis~\protect\cite{SPi0}
}
 \label{fig:sigpi0_pol_v17} 
\end{figure*}

 Our results for the $\Sigma^0$ polarization as a function
 of $\cos\theta^{\ast}(\pi^0)$ 
 for the $K^-p \to \pi^0 \Sigma^0$ reaction
 are given for each of the eight beam momenta in
 Tables~\ref{tab:spol1} and \ref{tab:spol2}.      
 These results are also shown in Fig.~\ref{fig:sigpi0_pol_v17}, compared
 to the VA analysis~\cite{SPi0}.
\begin{table*}
\caption
[tab:spol1]{
 Polarization of the $\Sigma^0$ hyperon in the $K^- p \to \pi^0 \Sigma^0$ reaction
 for the four lowest beam momenta.
 } \label{tab:spol1}
\begin{ruledtabular}
\begin{tabular}{|c|c|c|c|c|} 
\hline
 $p_{K^-}\pm \delta_p$ [MeV/$c$] & $514\pm 10$ & $560\pm 11$ &
                                     $581\pm 12$ & $629\pm 11$ \\
\hline
  $\cos\theta^{\ast}(\pi^0)$ & $P_{\Sigma^0}$ &
  $P_{\Sigma^0}$ & $P_{\Sigma^0}$ & $P_{\Sigma^0}$ \\
\hline
 -0.94 & $0.14\pm 0.25$ & $-0.28\pm 0.17$ & $-0.14\pm 0.14$ & $0.18\pm 0.15$ \\
 -0.81 & $0.32\pm 0.24$ & $-0.33\pm 0.17$ & $-0.10\pm 0.14$ & $-0.16\pm 0.13$ \\
 -0.69 & $-0.56\pm 0.24$ & $-0.09\pm 0.19$ & $-0.31\pm 0.15$ & $-0.11\pm 0.15$ \\
 -0.56 & $-0.77\pm 0.27$ & $-0.38\pm 0.19$ & $-0.28\pm 0.16$ & $-0.37\pm 0.16$ \\
 -0.44 & $-0.50\pm 0.35$ & $-0.69\pm 0.20$ & $-0.35\pm 0.18$ & $-0.37\pm 0.18$ \\
 -0.31 & $-0.39\pm 0.28$ & $-0.38\pm 0.22$ & $-0.27\pm 0.20$ & $-0.30\pm 0.19$ \\
 -0.19 & $-0.96\pm 0.31$ & $-0.53\pm 0.26$ & $-0.21\pm 0.19$ & $-0.18\pm 0.21$ \\
 -0.06 & $-0.77\pm 0.32$ & $-0.99\pm 0.26$ & $-0.59\pm 0.20$ & $-0.41\pm 0.21$ \\
 0.06 & $-1.02\pm 0.32$ & $-0.75\pm 0.26$ & $-0.71\pm 0.20$ & $-0.31\pm 0.21$ \\
 0.19 & $-0.75\pm 0.34$ & $-1.38\pm 0.29$ & $-0.73\pm 0.21$ & $-0.43\pm 0.22$ \\
 0.31 & $-0.16\pm 0.30$ & $-0.62\pm 0.24$ & $-0.69\pm 0.21$ & $-0.81\pm 0.21$ \\
 0.44 & $-1.37\pm 0.37$ & $-0.52\pm 0.25$ & $-0.44\pm 0.21$ & $-0.53\pm 0.21$ \\
 0.56 & $-0.37\pm 0.32$ & $-0.43\pm 0.28$ & $-0.87\pm 0.21$ & $-0.37\pm 0.23$ \\
 0.69 & $-0.14\pm 0.41$ & $-0.82\pm 0.31$ & $-0.50\pm 0.26$ & $-0.71\pm 0.25$ \\
 0.81 & $-0.84\pm 0.47$ & $-1.20\pm 0.43$ & $-0.38\pm 0.39$ & $0.32\pm 0.36$ \\
 0.94 & $-2.22\pm 0.84$ & $-0.75\pm 1.00$ &$1.99\pm 0.95$  & $-0.82\pm 0.69$ \\
\hline
\end{tabular}
\end{ruledtabular}
\end{table*}
\begin{table*}
\caption
[tab:spol2]{
 Polarization of the $\Sigma^0$ hyperon in the $K^- p \to \pi^0 \Sigma^0$ reaction
 for the four highest beam momenta.
 } \label{tab:spol2}
\begin{ruledtabular}
\begin{tabular}{|c|c|c|c|c|} 
\hline
 $p_{K^-}\pm \delta_p$ [MeV/$c$] & $659\pm 12$ & $687\pm 11$ &
                                     $714\pm 11$ & $750\pm 13$ \\
\hline
  $\cos\theta^{\ast}(\pi^0)$ & $P_{\Sigma^0}$ &
  $P_{\Sigma^0}$ & $P_{\Sigma^0}$ & $P_{\Sigma^0}$ \\
\hline
 -0.94 & $-0.18\pm 0.15$ & $-0.34\pm 0.12$ & $-0.09\pm 0.14$ & $0.03\pm 0.11$ \\
 -0.81 & $-0.22\pm 0.12$ & $-0.07\pm 0.11$ & $0.08\pm 0.11$ & $0.21\pm 0.10$ \\
 -0.69 & $-0.20\pm 0.13$ & $-0.07\pm 0.12$ & $-0.26\pm 0.13$ & $-0.03\pm 0.11$ \\
 -0.56 & $0.13\pm 0.15$ & $-0.28\pm 0.15$ & $-0.11\pm 0.14$ & $-0.01\pm 0.14$ \\
 -0.44 & $0.05\pm 0.18$ & $-0.35\pm 0.16$ & $0.16\pm 0.15$ & $-0.11\pm 0.16$ \\
 -0.31 & $-0.42\pm 0.18$ & $-0.11\pm 0.18$ & $-0.35\pm 0.19$ & $0.10\pm 0.17$ \\
 -0.19 & $-0.29\pm 0.19$ & $-0.53\pm 0.18$ & $-0.31\pm 0.20$ & $-0.39\pm 0.17$ \\
 -0.06 & $-0.20\pm 0.20$ & $-0.40\pm 0.20$ & $-0.03\pm 0.19$ & $0.18\pm 0.17$ \\
 0.06 & $-0.38\pm 0.20$ & $-0.79\pm 0.19$ & $-0.26\pm 0.19$ & $-0.03\pm 0.14$ \\
 0.19 & $-0.58\pm 0.19$ & $-0.41\pm 0.21$ & $-0.26\pm 0.18$ & $0.06\pm 0.13$ \\
 0.31 & $-0.50\pm 0.19$ & $-0.16\pm 0.18$ & $-0.20\pm 0.16$ & $0.20\pm 0.12$ \\
 0.44 & $-0.11\pm 0.19$ & $-0.63\pm 0.18$ & $-0.09\pm 0.17$ & $-0.03\pm 0.12$ \\
 0.56 & $-0.71\pm 0.21$ & $-0.18\pm 0.19$ & $-0.40\pm 0.18$ & $0.29\pm 0.12$ \\
 0.69 & $-0.06\pm 0.25$ & $-0.25\pm 0.22$ & $-0.08\pm 0.19$ & $0.14\pm 0.13$ \\
 0.81 & $-0.28\pm 0.36$ & $-0.42\pm 0.34$ & $-0.34\pm 0.32$ & $0.24\pm 0.24$ \\
 0.94 & --- & --- & --- & --- \\
\hline
\end{tabular}
\end{ruledtabular}
\end{table*}
 As seen, our results have smaller statistical uncertainties.
 Our sign for the polarization is opposite to the one from the VA analysis.
 Also, our magnitudes for the polarization at the lowest beam momenta are
 systematically smaller for the forward angles, in which the VA data
 have much smaller acceptance.
 We were not able to get any analysis details from
 the authors of Ref.~\cite{SPi0}
 that would help us to explain the difference between the two sets of results. 
 We want only to note that a similar measurement of the $\Lambda$ polarization
 gave us reasonable agreement with the data from Armenteros.

 The comparison of our results for
 the product of the $\Sigma^0$ polarization
 and the differential cross section of 
 $K^- p \to \pi^0 \Sigma^0$ with the data
 from Armenteros\cite{Arm70} is shown
 Fig.~\ref{fig:lampi0_pol_v17_arm}.
\begin{figure*}
\includegraphics[width=15.5cm,height=14.cm,bbllx=0.5cm,bblly=1.cm,bburx=19.5cm,bbury=18.5cm]{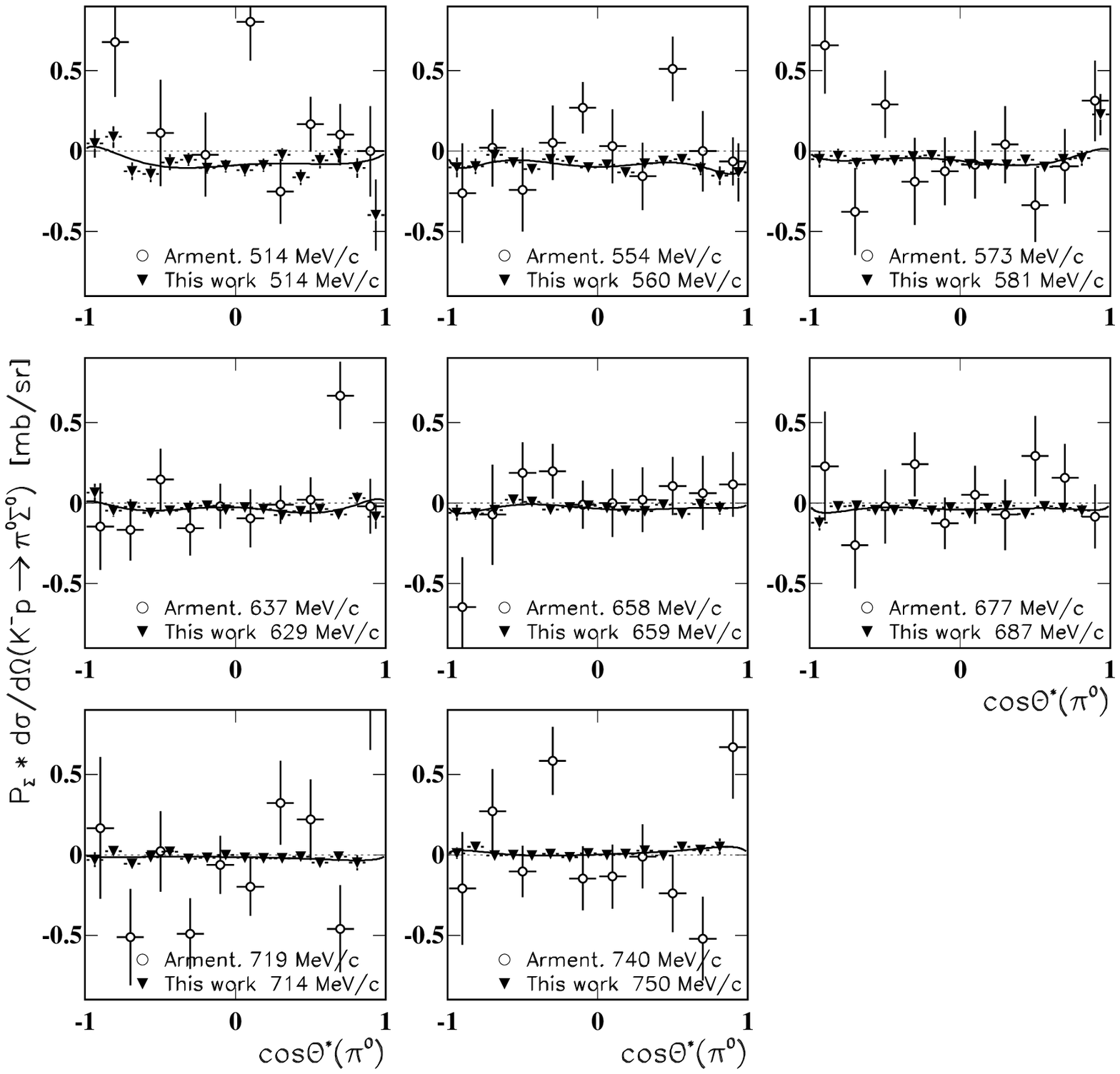}
\caption{
 Our results for the product of the $\Sigma^0$ polarization
 and the differential cross section of 
 $K^- p \to \pi^0 \Sigma^0$ compared to the data
 from Armenteros~\protect\cite{Arm70}.
 The curves are the fits of our data to
 the first associated Legendre functions. 
}
 \label{fig:sigpi0_pol_v17_arm} 
\end{figure*}
 The results from Armenteros
 have very large statistical uncertainties.
 Also, the quality of the data from
 Armenteros is poorer, as they could not measure
 $\Sigma^0$ and $\pi^0$ directly.
 The large uncertainties of the data from
 Armenteros do not allow the comparison of
 their sign of the $\Sigma^0$ polarization
 with ours.
 The curves in the figure are the fits
 of our $K^- p \to \pi^0 \Sigma^0$ data
 to the first associated Legendre functions.
 The results of the fits are given
 for each of the eight beam momenta in
 Tables~\ref{tab:ftspol1} and \ref{tab:ftspol2}.    
\begin{table*}
\caption
[tab:ftspol1]{
 Associated Legendre function coefficients for 
 the $K^- p \to \pi^0 \Sigma^0$ reaction for the four lowest beam momenta.
 } \label{tab:ftspol1}
\begin{ruledtabular}
\begin{tabular}{|c|c|c|c|c|} 
\hline
 $p_{K^-}\pm \delta_p$ [MeV/$c$] & $514\pm 10$ & $560\pm 11$ &
                                     $581\pm 12$ & $629\pm 11$ \\
\hline
 $B_1$ & $0.0895\pm 0.014$ & $0.098\pm 0.011$ & $0.0671\pm 0.0086$ & $0.0425\pm 0.0073$ \\
 $B_2$ & $0.007\pm 0.013$ & $0.009\pm 0.010$ & $0.0017\pm 0.0080$ & $0.0016\pm 0.0070$ \\
 $B_3$ & $-0.004\pm 0.012$ & $0.0191\pm 0.0097$ & $0.0012\pm 0.0077$ & $-0.0011\pm 0.0066$ \\
 $B_4$ & $0.0116\pm 0.0099$ & $0.0031\pm 0.0078$ & $-0.0117\pm 0.0062$ & $-0.0037\pm 0.0052$ \\
 $B_5$ & $-0.0050\pm 0.0095$ & $0.0147\pm 0.0071$ & $-0.0034\pm 0.0058$ & $-0.0091\pm 0.0049$ \\
 $\chi^2$/ndf & 1.64 & 0.78 & 0.75 & 0.93 \\
\hline
\end{tabular}
\end{ruledtabular}
\end{table*}
\begin{table*}
\caption
[tab:ftspol2]{
 Associated Legendre function coefficients for 
 the $K^- p \to \pi^0 \Sigma^0$ reaction for the four highest beam momenta.
 } \label{tab:ftspol2}
\begin{ruledtabular}
\begin{tabular}{|c|c|c|c|c|} 
\hline
 $p_{K^-}\pm \delta_p$ [MeV/$c$] & $659\pm 12$ & $687\pm 11$ &
                                     $714\pm 11$ & $750\pm 13$ \\
\hline
 $B_1$ & $0.0358\pm 0.0070$ & $0.0440\pm 0.0067$ & $0.0214\pm 0.0067$ & $-0.0137\pm 0.0055$ \\
 $B_2$ & $0.0024\pm 0.0069$ & $-0.0007\pm 0.0068$ & $0.0057\pm 0.0071$ & $-0.0096\pm 0.0060$ \\
 $B_3$ & $0.0079\pm 0.0067$ & $0.0087\pm 0.0066$ & $0.0057\pm 0.0071$ & $-0.0121\pm 0.0062$ \\
 $B_4$ & $-0.0073\pm 0.0052$ & $-0.0026\pm 0.0050$ & $0.0007\pm 0.0050$ & $0.0004\pm 0.0043$ \\
 $B_5$ & $0.0059\pm 0.0049$ & $0.0059\pm 0.0047$ & $0.0008\pm 0.0047$ & $-0.0028\pm 0.0037$ \\
 $\chi^2$/ndf & 1.24 & 1.52 & 1.15 & 1.20 \\
\hline
\end{tabular}
\end{ruledtabular}
\end{table*}

\section{Conclusions}

 Differential cross sections and hyperon
 polarizations have been measured
 for $\bar{K}^0 n$, $\pi^0 \Lambda$,
 and $\pi^0 \Sigma^0$ production in $K^- p$ interactions
 at eight $K^-$ momenta between 514 and 750~MeV/$c$.
 The experiment detected the multiphoton final
 states with the Crystal Ball spectrometer using
 a $K^-$ beam from the Alternating Gradient Synchrotron
 of BNL.
 The results provide significantly greater precision than
 the existing data, allowing a detailed reexamination
 of the excited hyperon states in our energy
 range.

\begin{acknowledgments}
 This work was supported in part by  DOE and NSF of the U.S.,
 NSERC of Canada, the Russian Ministry of Industry, 
 Science and Technologies, and the Russian Foundation for Basic Research.
 We thank SLAC for the loan of the Crystal Ball.
 The assistance of BNL and AGS with the setup is much appreciated.
\end{acknowledgments}

\end{document}